\documentclass[authoryear,12pt,letter,3p]{jowarticle}
\usepackage{graphicx}
\usepackage{amsmath}
\usepackage{amssymb}
\usepackage{setspace}
\usepackage{appendix}
\usepackage[all]{xy}
\makeatletter
\renewcommand\section{\@startsection{section}{1}{\z@}{-3.25ex plus -1ex minus -.2ex}{1.5ex plus .2ex}{\normalsize\bf}}
\renewcommand\subsection{\@startsection{subsection}{2}{\z@}{-3.25ex plus -1ex minus -.2ex}{1.5ex plus .2ex}{\normalsize\bf}}
\renewcommand\subsubsection{\@startsection{subsubsection}{3}{\z@}{-3.25ex plus -1ex minus -.2ex}{1.5ex plus .2ex}{\normalsize\bf}}
\makeatother

\begin{document}
\begin{frontmatter}
\title{Fiber Bundles, Yang-Mills Theory, and General Relativity}
\author{James Owen Weatherall}\ead{weatherj@uci.edu}
\address{Department of Logic and Philosophy of Science\\ University of California, Irvine, CA 92697}
\begin{abstract}I articulate and discuss a geometrical interpretation of Yang-Mills theory.  Analogies and disanalogies between Yang-Mills theory and general relativity are also considered.\end{abstract}
\end{frontmatter}

\doublespacing

\section{Introduction}

In recent philosophical discussions of (classical) Yang-Mills theory, a distinction is often made between ``holonomy interpretations'' of the theory and ``fiber bundle interpretations''.  Versions of the former have been carefully articulated and defended by, for instance, \citet{Belot1,Belot2}, \citet{Lyre}, and \citet{HealeyOldest,HealeyOld,Healey}.\footnote{\citet{Healey} is now (already) the \emph{locus classicus} for philosophy of Yang-Mills theory; the reader should refer there for an extensive discussion of the philosophical literature on the topic and for further references.}  The latter interpretation, meanwhile, is usually counted as the ``received view.''  But like many received views, it is not perfectly clear where it has been received from---or, for that matter, what it amounts to. Some oft-cited sources include \citet{Wu+Yang} and \citet{Trautman}, but these are written by and for working mathematicians and physicists, and neither takes on a distinctively ``interpretive'' posture.  More recently, some philosophers, such as \citet{Maudlin} and \citet{Arntzenius}, have \emph{endorsed} a fiber bundle interpretation---essentially by endorsing the fiber bundle formalism---and attempted to draw some philosophical morals regarding, for instance, classical property attribution.  But these authors appear to take for granted that the physical significance of the fiber bundle formalism for Yang-Mills theory is well understood.\footnote{One exception to this rule is \citet{Leeds}, who does attempt to articulate a novel fiber bundle interpretation of Yang-Mills theory, though the fiber bundle formalism he begins with is somewhat idiosyncratic with respect to the mathematical physics literature.  The view developed here is significantly different.  \citet{Catren}, too, offers an interpretation of Yang-Mills theory that makes extensive use of the fiber bundle formalism, but his goal is to relate this formalism to certain general ``principles'' that he detects in the foundations of Yang-Mills theory. This approach reflects a substantially different posture towards the foundations of physics from the one adopted here.  Perhaps the closest precursor to the present paper is \citet{HealeyOldest,Healey}, who describes and rejects a fiber bundle interpretation in the course of clearing the ground for his alternative holonomy interpretation.  But for reasons discussed in detail in section \ref{Disanalogies} of this paper, Healey rejects what I argue is the most natural interpretational strategy, leading him to a rather different view.}  And perhaps it \emph{is} well understood---certainly physicists have used the formalism successfully for decades.  Nonetheless, it is not clear that anyone has succeeded in articulating just what that physical significance is supposed to be, at least not in a way that shows how the geometry of principal bundles leads to a fiber bundle \emph{interpretation} of Yang-Mills theory.  Doing so is the principal goal of the present paper---though, as I will discuss below, there remains work to do.\footnote{I will \emph{not} discuss the relative merits of fiber bundle and holonomy interpretations of Yang-Mills theories.  (For more on that topic, see \citet{Rosenstock+Weatherall1, Rosenstock+Weatherall2}.)  And although I do not know of anyone who has written a philosophical treatment along the lines of what follows, I do not claim that the views presented here are original.  Indeed, much of what I say is implicit in classical sources in the mathematical physics literature, especially \citet{Palais} and \citet{Bleecker}.  That said, in both sources the themes emphasized here are somewhat obscured by the authors' treatments of Kaluza-Klein theory, which puts the same formalism to strikingly different physical uses.}

My basic strategy will be to exploit an analogy between Yang-Mills theory and general relativity, where I take it we are on firmer ground.\footnote{Of course, how one should understand general relativity is itself a matter of some continuing controversy (see \citet{Brown}); nonetheless, it seems we are still on \emph{firmer} ground than in Yang-Mills theory.  As will be clear in what follows, I take general relativity to be a theory of spatiotemporal geometry, along the lines of what is described in, say, \citet{Wald} or \citet{MalamentGR}.}  The mathematical facts on which the analogy relies are (mostly) well known: general relativity, like Yang-Mills theory, may be conceived as a theory of a principal connection on a certain principal bundle over spacetime.  Usually, this fact is cited in connection with attempts to translate general relativity into the language of Yang-Mills theory, as a first step towards generalizing or quantizing general relativity.  Here, however, I will use the analogy to translate in the opposite direction.  The idea is that, if one is accustomed to thinking of general relativity as a theory of spacetime geometry, then there is a natural way of understanding the principal bundle on which the reformulation noted above is based.  More importantly, there is a straightforward, largely deflationary, interpretation of so-called ``gauge transformations'' and ``gauge invariance'' in general relativity.  Drawing on the analogy between Yang-Mills theory and general relativity, then, I will argue that one can understand the geometry of Yang-Mills theory in precisely the same way as in general relativity.  In other words, though one often sees claims that general relativity may be cast in a form that makes it strongly analogous to Yang-Mills theory, I will argue that so-too can Yang-Mills theory be cast so as to make it strongly analogous to general relativity.  An important upshot of this argument will be that the deflationary interpretation of the significance of ``gauge transformations'', ``gauge invariance'', and related notions carries over to Yang-Mills theory.

The picture that emerges is one on which some matter has local degrees of freedom---states and properties---that are best conceived as ``vectorial'', in the sense that one can make sense of adding them or multiplying them by scalars.    Familiar examples of vectorial properties include electromagnetic charge or velocity, and indeed, one often encounters matter with vectorial properties in general relativity.  To represent such matter on spacetime, it is convenient to use smooth vector (and tensor) fields to encode the distribution of that matter in space and time.  But not all such properties are naturally represented by ``ordinary'' vector fields on spacetime---that is, by sections of the tangent bundle---since not all such properties are naturally understood to correspond to tangents to smooth curves in spacetime.  So to represent such properties, we must introduce different kinds of vector fields on spacetime.  These are understood as sections of vector bundles over spacetime.  The basic insight of Yang-Mills theory, then, is that the derivative operators acting on these vector bundles are curved---and that the curvature of these bundles is related to the distribution of (certain kinds) of matter in space and time, just as in general relativity the curvature of spacetime is related to the distribution of energy-momentum in space and time.  Influences that we would have otherwise called ``forces'', such as the electromagnetic force or the strong force, then, are manifestations of the curvature of these bundles.

On this picture, the principal bundles and principal connections that one often associates with Yang-Mills theory play an auxiliary role: they arise as a way of coordinating derivative operators acting on different, but systematically related, vector bundles associated with different kinds of matter influenced by the same forces.  Choices of gauge, meanwhile, are nothing more (or less) than choices of frame field (i.e., bases) for these vector bundles, relative to which one may conveniently represent the derivative operators for certain calculations; gauge invariance, meanwhile, is simply the requirement that properties attributed to such fields do not depend on a particular choice of basis.

I will develop this view in detail in what follows.  But first, let me flag what I take to be the most significant shortcoming of the picture just sketched---and also what needs to be done to finish the present project.  As I develop it here, the significance of the principal bundle formalism for Yang-Mills theory is given by the role that the principal connection plays in the dynamics of matter, represented by sections of some vector bundle.  But there is something immediately unsatisfactory about this approach, which is that the ``classical'' matter fields one encounters here---complex scalar fields, say, or classical ``quark'' fields---do not appear to represent any realistic physical systems.\footnote{One possible exception occurs in theories of cosmological inflation, where the ``inflaton'' field is often taken to be a smooth scalar field satisfying some non-linear wave equation.  But since this field is not charged, its dynamics would not depend on a principal connection.}  Thus the ``vectorial properties'' associated with such fields are (at best) abstract.

This problem, or something closely related, plagues \emph{any} discussion of classical Yang-Mills theory, but it is cast in particularly stark relief by what I say here.  It seems to me that to give a completely satisfactory account of the significance of the principal bundle formalism, one needs to say more than I am able to say here about the physical significance of classical matter fields.  One option is to understand these fields as a way of encoding energy-momentum and charge-current density properties of some (unrealistic) matter.  But if this is right, it would be desirable to have a way of thinking about classical Yang-Mills theory that dealt only in terms of these concrete properties---perhaps by characterizing how a connection on a principal bundle influences the motion of appropriately charged test particles, or directly determines the dynamics of energy-momentum and charge-current density fields.  In any case, filling this gap is an important task that I do not attempt here, and for that reason the present paper may best be construed as a prolegomena to a full account of the physical significance of the principal bundle formalism.  

The remainder of the paper will be organized as follows.  I will begin by giving an overview of Yang-Mills theory from a geometrical perspective.  Here I simply follow standard treatments in the mathematics and mathematical physics literature; I will say little concerning what this geometry represents physically.  Next I will describe how general relativity may be reformulated in the language of principal bundles, emphasizing how to understand this new geometrical setting for the theory in light of more familiar presentations.  I will then return to Yang-Mills theory and show how one can import the interpretation of the formalism one gets from general relativity to that context.  Finally, I will discuss some limitations of the analogy between general relativity and Yang-Mills theory, including some that have been observed previously in the literature, most notably by \citet{Trautman}, \citet{Anandan}, and \citet{HealeyOld, Healey}.  I will argue that, while important disanalogies certainly exist, their significance for geometrical interpretations of Yang-Mills theory have been over-stated.  Finally, I include an appendix that presents the formalism of principal bundles, vector bundles, and connections in a language that would be familiar to readers of, say, \citet{Wald} or \citet{MalamentGR}.  This appendix is intended both to orient readers to the notation used throughout the present paper and also to serve as an invitation to philosophers of physics trained in the foundations of spacetime theories to further investigate the geometry underlying Yang-Mills theory.

\section{The Geometry of Yang-Mills Theory}\label{Yang-Mills}

A model of Yang-Mills theory consists in (1) a \emph{relativistic spacetime} $(M,g_{ab})$,\footnote{I am taking for granted, here, as in the appendix, some familiarity with both the mathematics and the physics of general relativity.  For sympathetic treatments of both topics, see \citet{Wald} and \citet{MalamentGR}.} which is a smooth four dimensional manifold with a Lorentz-signature metric,\footnote{All manifolds considered here are assumed to be Hausdorff, paracompact, and smooth; more generally, all maps and fields that are candidates to be smooth will also be assumed to be smooth.} (2) a principal bundle $G\rightarrow P\xrightarrow{\wp} M$ over the spacetime manifold with structure group $G$,\footnote{\label{AI} The question of what postulating such a structure means physically will be the primary concern of this paper, and will be addressed in subsequent sections. For now, we just take for granted that this \emph{is} the geometrical setting in which we are working.  For mathematical background, see Appendix \ref{Preliminaries} and references therein.  One notational convention is worth mentioning here, however: throughout, I use the ``abstract index'' notation developed by \citet{Penrose+Rindler} and described in detail by \citet{Wald} and \citet{MalamentGR}, suitably modified to distinguish the range of vector spaces that one encounters in the theory of principal bundles.  A detailed discussion of these modifications is given in \ref{AppNotation}.} and (3) an invariant inner product $k_{\mathfrak{A}\mathfrak{B}}$ on the Lie algebra $\mathfrak{g}$ associated to $G$.\footnote{The invariance of the inner product $k_{\mathfrak{A}\mathfrak{B}}$ is with respect to the adjoint action of $\mathfrak{g}$ on itself.  In the case of a compact Lie group, there is an essentially unique choice of inner product up to scaling factor.  This scaling factor is closely related to the ``coupling constant'' associated with a Yang-Mills theory, i.e., it provides a measure of the relative ``strength of interaction'' of different Yang-Mills fields.  The inner product is necessary to define, for instance, energy-momentum tensors and Lagrangians for Yang-Mills fields, but it plays no role in the present paper.}  The bundle $P$ is assumed to be endowed with (4) a principal connection, $\omega^{\mathfrak{A}}{}_{\alpha}$, which is a smooth Lie algebra valued one form on $P$ that maps tangent vectors $\xi^{\alpha}$ at a point $x\in P$ to elements of the Lie algebra $\mathfrak{g}$.  Associated with the connection is a curvature,
\begin{equation}\label{curvature}
\Omega^{\mathfrak{A}}{}_{\alpha\beta} = d_{\alpha}\omega^{\mathfrak{A}}{}_{\beta} + \frac{1}{2}[\omega^{\mathfrak{A}}{}_{\alpha},\omega^{\mathfrak{A}}{}_{\beta}].
\end{equation}
Here $d_{\alpha}$ is the exterior derivative and the bracket is the Lie bracket on $\mathfrak{g}$.\footnote{To be clear about this notation, the curvature is the tensor that takes vectors $\xi^{\alpha},\eta^{\alpha}$ at a point $x\in P$ to the Lie algebra element $\xi^{\alpha}\eta^{\beta}d_{\alpha}\omega^{\mathfrak{A}}{}_{\beta} + \frac{1}{2}[\omega^{\mathfrak{A}}{}_{\alpha}\xi^{\alpha},\omega^{\mathfrak{A}}{}_{\beta}\eta^{\beta}]$.  See \ref{AppCurvature}.}  The connection $\omega^{\mathfrak{A}}{}_{\alpha}$ is sometimes called the ``Yang-Mills field'', while the curvature is called the ``field strength''.  A model of the theory might then be written $(P,\omega^{\mathfrak{A}}{}_{\alpha},g_{ab})$, where we use $P$ to abbreviate the entire principal bundle structure, including the base manifold and structure group.

Suppose we have a model of Yang-Mills theory, $(P,\omega^{\mathfrak{A}}{}_{\alpha},g_{ab})$.  The fundamental dynamical principal of Yang-Mills theory, the Yang-Mills equation, may be written as:
\begin{equation}\label{YME}
\star\overset{\omega}{D}{}_{\alpha}\star\Omega^{\mathfrak{A}}{}_{\beta\kappa}=J^{\mathfrak{A}}{}_{\kappa}.
\end{equation}
Here $\overset{\omega}{D}$ is the exterior covariant derivative relative to $\omega^{\mathfrak{A}}{}_{\alpha}$ and $\star$ is a Hodge star operator defined relative to $\omega^{\mathfrak{A}}{}_{\alpha}$ and $g_{ab}$.\footnote{See \ref{AppExterior+Induced} and \ref{AppHodgeStar}, respectively, for more on exterior covariant derivatives and this Hodge star.  Note that, although the Hodge star notation is a somewhat unnatural fit with the index notation, Eq. \eqref{YME} makes sense: $\star\Omega^{\mathfrak{A}}{}_{\beta\kappa}$ is a vector valued two form, so $\overset{\omega}{D}{}_{\alpha}\star\Omega^{\mathfrak{A}}{}_{\beta\kappa}$ is a vector valued three form, and thus $\star\overset{\omega}{D}{}_{\alpha}\star\Omega^{\mathfrak{A}}{}_{\beta\kappa}$ is a vector valued one form.}  The field $J^{\mathfrak{A}}{}_{\alpha}$, meanwhile, is a horizontal and equivariant one form on $P$, representing the total Yang-Mills \emph{charge-current density} associated with any matter present.  It acts as a source term in the Yang-Mills equation.

The fields $\omega^{\mathfrak{A}}{}_{\alpha}$, $\Omega^{\mathfrak{A}}{}_{\alpha\beta}$, and $J^{\mathfrak{A}}{}_{\alpha}$ are all defined on the total space of the principal bundle, $P$.  It is frequently convenient, however, to think of them instead as fields on spacetime, in order to study how such fields would interact with other fields on spacetime or with various measurement apparatuses, which are also represented by fields on spacetime.  Unfortunately, there is no general, canonical way to represent fields on $P$ as fields on $M$.  What one \emph{can} do is first choose a local section $\sigma:U\rightarrow P$ of $P$, where $U\subseteq M$ is open, and then represent the fields by pulling them back to $M$ along $\sigma$, as $\sigma^*(\omega^{\mathfrak{A}}{}_{\alpha})$, $\sigma^*(\Omega^{\mathfrak{A}}{}_{\alpha\beta})$, and $\sigma^*(J^{\mathfrak{A}}{}_{\alpha})$.\footnote{For more on the pullback of vector valued forms, see \ref{AppVectorValuedForms}.} It is important to emphasize, however, that these representations generally depend on the choice of section.

Choosing a section relative to which to represent the Yang-Mills connection, curvature, and charge-current density on spacetime is sometimes called choosing a ``gauge''; changing from one choice of section to another is called a ``gauge transformation''.  In general, any change from one section $\sigma:U\rightarrow P$ to another $\sigma':U\rightarrow P$, where we suppose for convenience that both sections have the same domain, may be understood to correspond to a vertical principal bundle automorphism $f:P\rightarrow P$ such that $\sigma'=\sigma\circ f$.\footnote{Vertical principal bundle automorphisms are defined in \ref{AppVector+PrincipalBundles}.}  Conversely, any vertical principal bundle automorphism implements a change of section in this way.  Thus we may characterize the possible gauge transformations by characterizing the vertical principal bundle automorphisms of a given principal bundle.  This correspondence is useful because the automorphisms of a principal bundle may be understood in terms of the right action of the group $G$ on $P$, which in turn allows us to relate the changes in our representations of the dynamical fields on spacetime arising from changes of section to actions of the structure group on those fields.\footnote{One has to be slightly careful: the right action on $P$ by a \emph{fixed} element $g\in G$ gives rise to a vertical principal bundle automorphism only when $G$ is Abelian; more generally, the $G$ action will vary from point to point.  See \citet[\S3.2]{Bleecker} for a lucid and complete treatment of the vertical principal bundle automorphisms of a principal bundle $P$.} Note that we do not require these automorphisms to preserve the connection $\omega^{\mathfrak{A}}{}_{\alpha}$, so it would be misleading to think of these as ``symmetry transformations'' of models of Yang-Mills theory.

The theory I have described so far might be thought of as ``pure'' Yang-Mills theory, insofar as I have only described the Yang-Mills fields and their dynamics.  (Of course, I have included a source term in Eq. \eqref{YME}, but I have said nothing about the fields that would contribute to that source term.)  One is generally interested in the ways in which Yang-Mills fields interact with other fields.  So-called ``matter fields'' that interact with Yang-Mills fields are thought of as sections of vector bundles $V\rightarrow P\times_{G} V\xrightarrow{\pi_{\rho}} M$ over $M$ associated to $P$.\footnote{Once again, for further details on the associated bundle construction, see \ref{AppAssociatedBundles}.}  Matter is then represented by sections of this vector bundle over spacetime---i.e., as (generalized) vector fields on (open subsets of) $M$.  (These, in turn, may be associated with charge-current densities $J^{\mathfrak{A}}{}_{\alpha}$ on $P$, which contribute to the Yang-Mills equation, as well as energy-momentum tensors on $M$.)  The principal connection $\omega^{\mathfrak{A}}{}_{\alpha}$ on $P$ induces a covariant derivative operator $\overset{\omega}{\nabla}$ on $P\times_{G} V$, which then appears as the standard of differentiation in the field equations governing these matter fields.  Thus, solutions to the Yang-Mills equation influence the evolution of matter fields by determining the derivative operator appearing in the dynamics of those fields; conversely, matter fields give rise to charge-current densities that serve as source terms in the Yang-Mills equation.

This way of presenting Yang-Mills theory focuses on its geometrical features, which will be our central concern in subsequent sections.  But it is fairly abstract, so it may be useful to recall how this formalism relates to concrete examples.  The simplest, and most familiar, example of a Yang-Mills theory is classical electromagnetism, which we will briefly develop here.  We will then mention some other examples, though we will not develop them in any detail.

The structure group for electromagnetism is $U(1)$, the unitary group of degree 1, which may be thought of as the circle, $\mathcal{S}^1$, endowed with a group structure by recognizing that $\mathcal{S}^1$ can be embedded in $\mathbb{C}$ (the complex plane, understood as a two dimensional real manifold) in such a way that the image of $\mathcal{S}^1$ is precisely the complex numbers of unit norm.  The complex numbers of unit norm, meanwhile, have a natural group structure given by addition of their complex phase mod $2\pi$.  The group $U(1)$ is Abelian, i.e., commutative, and its associated Lie algebra, $\mathfrak{u}(1)$, is isomorphic to $\mathbb{R}$, since $\mathcal{S}^1$ is a one dimensional manifold.  The Lie bracket is trivial, as required for the Lie algebra of an Abelian Lie group: given any $a,b\in\mathbb{R}$, $[a,b]=ab-ba=0$.  The inner product on $\mathfrak{u}(1)$, meanwhile, is just ordinary multiplication.\footnote{Up to a choice of scaling factor.}

Thus, a model of electromagnetism consists of a spacetime $(M,g_{ab})$, a $U(1)$ bundle $U(1)\rightarrow EM\xrightarrow{\wp} M$ over $M$, and a principal connection $\omega^{\mathfrak{A}}{}_{\alpha}$ on $P$.  In this case, however, since $\mathfrak{u}(1)$ is just $\mathbb{R}$, $\omega^{\mathfrak{A}}{}_{\alpha}$ is a real valued one form, i.e., an ordinary linear functional.  Thus we may freely drop the $\mathfrak{A}$ index altogether and write $\omega_{\alpha}$ for the connection.  Similarly, the curvature may now be written as:
\begin{equation}\label{EMcurvature}
\Omega_{\alpha\beta} = d_{\alpha}\omega_{\beta} + \frac{1}{2}[\omega_{\alpha},\omega_{\beta}]=d_{\alpha}\omega_{\beta},
\end{equation}
where the second equality follows because the Lie algebra is Abelian.

To recover more traditional presentations of electromagnetism, we choose a local section $\sigma:U\rightarrow EM$.  Relative to this section, we define fields $A_a=\sigma^*(\omega_{\alpha})$, $F_{ab}=\sigma^*(\Omega_{\alpha\beta})$, and $J_a=\sigma^*(J_{\alpha})$.  Then, since the exterior derivative commutes with pullbacks, we see that:
\begin{equation}\label{E&M}
d_aA_b = d_a\sigma^*(\omega_{\beta})=\sigma^*(d_{\alpha}\omega_{\beta})=\sigma^*(\Omega_{\alpha\beta})=F_{ab}.
\end{equation}
We thus recognize $A_a$, the local representative of $\omega_{\alpha}$ relative to some local section $\sigma$, as what would otherwise be called a vector potential, and $F_{ab}$ as the associated electromagnetic field.  Note that because $U(1)$ is Abelian, horizontal and equivariant Lie algebra valued forms, such as $\Omega_{\alpha\beta}$ and $J_{\alpha}$, are invariant under vertical principal bundle automorphisms. Thus $F_{ab}$ and $J_a$ are independent of the choice of section.  It follows that if we chose a different section, $\sigma':U\rightarrow EM$, we would find that $\sigma^*(\omega_{\alpha})-\sigma'^*(\omega_{\alpha})$ is a closed one form, and thus, at least locally, it must be exact---i.e., locally there exists a scalar field $\psi$ such that $\sigma^*(\omega_{\alpha})=\sigma'^*(\omega_{\alpha}) + \nabla_a\psi$.  Thus we see that a change of section relative to which we represent $\omega_{\alpha}$ on spacetime yields a ``gauge transformation'' in the more traditional sense.

It follows from Eq. \eqref{E&M} that $F_{ab}$ is antisymmetric (by the definition of the exterior derivative) and that $d_aF_{bc}=\mathbf{0}$, since $F_{ab}$ is exact.  Thus we recover one of Maxwell's equations from purely geometrical considerations.  We recover the second of Maxwell's equations, meanwhile, by pulling back both sides of the Yang-Mills equation, Eq. \eqref{YME}, along $\sigma$:
\begin{equation}\label{Maxwell}
\sigma^*(\star \overset{\omega}{D}_{\alpha}\star\Omega_{\beta\kappa})=\sigma^*(\star d_{\alpha}\star\Omega_{\beta\kappa})=\star d_{a} \star \sigma^*(\Omega_{\beta\kappa})=\star d_a\star F_{bc}=\nabla_n F^n{}_c=J_c,
\end{equation}
Here we have used the facts that both $\star$ and the exterior derivative commute with pullbacks and that $U(1)$ is Abelian, which means the exterior covariant derivative reduces to the ordinary exterior derivative on horizontal and equivariant vector valued $k$ forms.  The final equality in Eq. \eqref{Maxwell} is precisely the inhomogeneous Maxwell equation in the ordinary form.

We now consider how electromagnetism interacts with matter, again in the simplest case.\footnote{\label{dis2} Let me once again echo the dissatisfaction noted above.  The complex scalar field described here, though a useful toy example, does not appear to represent any realistic matter.}  Take a one dimensional complex vector space $V$ and a faithful representation $\rho:U(1)\rightarrow GL(V)$ of $U(1)$ on this vector space.\footnote{Note that we might have equally well begun with a two dimensional real vector space. This will be important in section \ref{MainEvent}.}  These choices yield a vector bundle $V\rightarrow EM\times_{U(1)} V\xrightarrow{\pi_{\rho}} M$ over $M$ associated to $EM$.  A section $\varphi:U\rightarrow EM\times_{U(1)} V$ of this bundle represent states of a ``complex scalar field'' on spacetime, which is often taken as a representation of a simple kind of charged matter.

The connection $\omega_{\alpha}$ on $EM$ induces a covariant derivative operator $\overset{\omega}{\nabla}$ on this vector bundle, which allows one to, for instance, parallel transport field values along curves.    Similarly, the covariant derivative operator allows one to consider dynamics for sections $\varphi:U\rightarrow EM\times_{U(1)} V$ of the associated bundle.  A standard example is the Klein-Gordon equation, $\overset{\omega}{\nabla}_a\overset{\omega}{\nabla}{}^a\varphi=m^2\varphi$, where $m$ is a ``mass'' parameter.  One may also define energy momentum and charge-current density tensors.  For instance, for fields satisfying the Klein-Gordon equation, the energy-momentum tensor is given by $T_{ab}=\nabla_a\varphi\nabla_b\varphi + (\nabla_a\varphi)^*(\nabla_b\varphi)^*-g_{ab}\nabla_a\varphi(\nabla^a\varphi)^* + g_{ab}m^2\varphi^*\varphi$ and the charge-current density is $J_a=\varphi(\nabla_a\varphi)^*-\varphi^*\nabla_a\varphi$.\footnote{Note that in this case, we may define the charge-current density as a field on $M$ because, as noted above, $J_a$ is independent of the choice of section of $EM$.  In a more general setting, we would define $J^{\mathfrak{A}}{}_{\alpha}$ as a field on the total space of the principal bundle.}  Note that we may rewrite the last several equations in a (perhaps) more familiar form by noting that any choice of section $\sigma:U\rightarrow EM$ of the principal bundle gives rise to a corresponding flat derivative operator $\partial$ on sections of $EM\times_{U(1)} V$.  The action of the covariant derivative operator may then be written in terms of $\partial$ and $\sigma^*(\omega_{\alpha})=A_a$, so that for any $\varphi:U\rightarrow EM\times_{U(1)} V$, $\overset{\omega}{\nabla}\varphi = (\partial_a - iA_a)\varphi$.  %It is natural to think of $\partial$ and $A_a$ here as analogous to a coordinate derivative operator and associated Christoffel symbols.

Other physically important examples of Yang-Mills theories include the theories of the electroweak force, the strong force, and various ``grand unified theories''; these correspond to principal bundles with structure groups $SU(2)\times U(1)$, $SU(3)$, and (usually) $SU(n)$ or $SO(n)$ for larger $n$.  (Here $SU(n)$ refers to the special unitary group of degree $n$, which is isomorphic to the group of unitary $n\times n$ matrices with determinant 1; $SO(n)$, meanwhile, is the special orthogonal group of degree $n$, which is isomorphic to the group of orthogonal $n\times n$ matrices with determinant $1$.)  Matter represented by sections of vector bundles associated to these principal bundles include quark and neutrino fields.  These examples differ from electromagnetism in that their structure groups are generally non-Abelian, which has a number of consequences.  For instance, the relationship between the principal connection and curvature given in Eq. \eqref{curvature} does not simplify as in Eq. \eqref{EMcurvature}; likewise, the form of the Yang-Mills equation relative to a choice of section of the relevant principal bundle, given for electromagnetism in Eq. \eqref{Maxwell}, becomes more complicated.  Another important difference is that in the non-Abelian case, horizontal and equivariant Lie algebra valued forms on a principal bundle are not invariant under vertical principal bundle automorphisms.  Thus, unlike in electromagnetism, the field strength and charge-current density associated with other Yang-Mills theories will generally depend on the choice of section used to represent them as fields on spacetime.

\section{General Relativity and the Frame Bundle}\label{GR}

The Yang-Mills theories just described are undoubtedly geometrical---after all, they are theories of connections on a certain class of manifold.  But at first pass, they seem quite different from the most familiar---at least among philosophers of physics---geometrical theories in physics. In particular, general relativity and its cousins, such as the geometrized formulation of Newtonian gravitation,\footnote{See, for instance, \citet{TrautmanNC} and \citet[Ch. 4]{MalamentGR}.} are theories of \emph{spatio-temporal} geometry, in the sense that the theories represent certain geometrical properties that we ascribe to space and time.\footnote{Another class of physical theories that philosophers have studied recently---see, for instance, \citet{Belot}, \citet{Butterfield}, \citet{North}, \citet{Curiel}, and \citet{Barrett}---that make significant use of the sorts of geometrical methods discussed here are Hamiltonian and Lagrangian mechanics.  In that context one works with the cotangent and tangent bundles, respectively, of the manifold of possible configurations of some physical system.  As I hope will be clear in what follows, although similar methods are used, there are no strong analogies between this application of geometry in physics and Yang-Mills theory.  In particular, nothing in Yang-Mills theory as I have described it here should be understood as a manifold whose points are (global) configurations or instantaneous states of any physical system.}   Models of general relativity---relativistic spacetimes---are manifolds endowed with a Lorentz-signature metric, $(M,g_{ab})$, where the points of the manifold represent events in space and time and the metric represents facts about spatial distance and duration as determined by various observers.  The metric also classifies tangent vectors as spacelike, timelike, or null, and by extension allows us to classify curves and hypersurfaces as spacelike, timelike, or null.  Timelike curves are possible trajectories for idealized observers and for small (point-like) massive bodies; light in vacua follows null geodesics. The metric gives rise to a (unique) torsion-free covariant derivative operator $\nabla$ satisfying $\nabla_a g_{bc}=\mathbf{0}$, known as the Levi-Civita derivative operator.  This derivative operator provides a standard of geodesy, or non-acceleration, for objects in spacetime, and so in the absence of external forces, we understand small bodies to follow timelike geodesics.  And so on.

As we have seen, many of these same geometrical objects---albeit with more bells and whistles, and in greater generality---appear in the context of Yang-Mills theory.  There, too, we consider manifolds, covariant derivative operators, (generalized) fields on spacetime, etc.  But in a sense, Yang-Mills theory tacks all of this on as \emph{additional} geometry on top of the spatio-temporal geometry of relativity theory.  After all, a model of Yang-Mills theory begins with a relativistic spacetime, with its usual interpretation, and then adds a new, substantially larger manifold (the total space $P$), as well as a family of systematically related manifolds (the associated bundles).  If points in the spacetime manifold represent events in spacetime, what do points in the total space of a principal bundle over spacetime represent?  Presumably they do not represent additional events, related in some non-spatio-temporal fashion to ordinary events---or if they do, a great deal needs to be said about what the significance of those events is meant to be.

We have said previously that sections of a bundle may be thought of as (generalized) fields, and indeed, we have said that sections of associated vector bundles may represent certain distributions of matter (or distributions of properties of matter) on spacetime.  Vector bundles, then, at least in some cases, might be thought of as representing the possible field states at each point of spacetime, in the sense that the fiber over each point represents possible local states of some physical system.  Can one think of a principal bundle in the same way?  Thus far, sections of principal bundles have served only as representational devices: they have allowed us to represent fields on the total space as fields on spacetime.  But nothing has been said about the significance of the points in the bundle in the first place, or of what the relationship is supposed to be between the Yang-Mills field and field strength---i.e., the connection and curvature on the principal bundle---and the matter fields represented by sections of an associated bundle.

It is in connection with these questions that general relativity can provide significant insight into Yang-Mills theory, and even, I claim, provide the starting point for a ``fiber bundle interpretation'' (or, perhaps better, an interpretation of the fiber bundles).  The reason is that, as I noted above, general relativity, too, may be conceived as a theory of a connection on a principal bundle and induced derivative operators on associated bundles.  Thinking of general relativity in this way provides a different perspective on the role of principal bundles in Yang-Mills theory.

The first step in understanding general relativity as a theory of a connection on a principal bundle is to identify the relevant principal bundle.  To do so, we observe that, associated with \emph{any} $n$ dimensional manifold $M$, there is a canonical principal bundle known as the frame bundle, $GL(n,\mathbb{R})\rightarrow LM \xrightarrow{\wp_L} M$.\footnote{Some more details on the construction of the frame bundle are provided in \ref{AppTangent+FrameBundles}.} (Here $GL(n,\mathbb{R})$ is the general linear group of degree $n$, which is isomorphic to the group of invertible $n\times n$ matrices.)  The frame bundle over a manifold $M$ is the bundle whose fiber at each point $p$ consists in all (ordered) bases---or, ``frames''---for $T_pM$, the tangent space at $p$.\footnote{Generally in what follows, when we refer to ``bases'', we mean ``ordered bases''.}  %Since any two bases of $T_p M$ for any point $p\in M$ are related by some unique automorphism of $T_pM$,\footnote{In other words, given any two bases $u=\{\overset{1}{u}{}^A,\ldots\overset{n}{u}{}^A\}, v=\{\overset{1}{v}{}^A,\ldots\overset{n}{v}{}^A\}$ for any $n$-dimensional vector space (written relative to some fixed basis), there exists a unique invertible matrix $g^A{}_B$---i.e., a unique element $g\in GL(n,\mathbb{R})$---such that $g^A{}_B$ maps $u$ to $v$, in the sense that $\overset{i}{v}{}^A=g^A{}_B\overset{i}{u}{}^B$ for all $i$.} one can construct diffeomorphisms from the fibers of $LM$ to $GL(n,\mathbb{R})$ by sending some (any) $u\in (\pi_L)^{-1}[p]$ to the identity in $GL(,\mathbb{R})$, and then sending every other frame  $v\in(\pi_l)^{-1}[p]$ to the unique element of $GL(n,\mathbb{R})$ taking $u$ to $v$.  Similarly, there is a natural right action of $GL(n,\mathbb{R})$ on $LM$, corresponding to changing frame at a point.

As with any principal bundle, we may also consider vector bundles associated with the frame bundle.  In particular, an $n$ dimensional vector space $V$ with a faithful representation of $GL(n,\mathbb{R})$ gives rise to a vector bundle $V\rightarrow LM\times_{GL} V\xrightarrow{\pi_{\rho}} M$ over $M$ associated to $LM$.  Any such vector bundle is isomorphic to the tangent bundle of $M$, $V\rightarrow TM\xrightarrow{\pi_T} M$, which is the vector bundle whose fiber at each point is the tangent space at that point.\footnote{Again, for details of the construction of the tangent bundle, see \ref{AppTangent+FrameBundles}.}  In this sense, one may think of the tangent bundle as associated to the frame bundle; similarly, the cotangent bundle and bundles of rank $(r,s)$ tensors may be thought of as associated to the frame bundle.  As with any associated bundle, a principal connection $\omega^{\mathfrak{A}}{}_{\alpha}$ on $LM$ induces a unique covariant derivative operator $\nabla$ on $TM$ and the corresponding bundles of covectors and tensors; conversely, given any covariant derivative operator $\nabla$ on $TM$, there is a unique principal connection $\omega^{\mathfrak{A}}{}_{\alpha}$ on $LM$ that induces $\nabla$.

We are now in a position to describe general relativity in the language of principal bundles.  The translation manual is strikingly simple.  Fix a relativistic spacetime $(M,g_{ab})$ and consider the frame bundle $GL(n,\mathbb{R})\rightarrow LM\xrightarrow{\wp_L} M$ over the spacetime manifold $M$.  Fix an isomorphism $\vartheta$ between an appropriate associated vector bundle and the tangent bundle.  Then there is a unique principal connection $\omega^{\mathfrak{A}}{}_{\alpha}$ on $LM$ such that the induced derivative operator on $TM$ is torsion-free and satisfies $\nabla_ag_{bc}=\mathbf{0}$.  We require that the Riemann curvature tensor associated with $\nabla$, $R^a{}_{bcd}$, satisfy Einstein's equation, $R_{ab}-\frac{1}{2}g_{ab} R= T_{ab}$, for some energy-momentum tensor field $T_{ab}$, where $R_{ab}=R^a{}_{bca}$ and $R=g^{ab}R_{ab}$.  All of these fields are to be understood as sections of the appropriate bundles, which in turn are vector bundles associated to $LM$.  The Riemann curvature tensor, meanwhile, may be re-expressed using the curvature $\Omega^{\mathfrak{A}}{}_{\alpha \beta}$ on $LM$ associated with $\omega^{\mathfrak{A}}{}_{\alpha}$, so that Einstein's equation may be understood as governing the curvature of the principal connection.  In this sense, general relativity is a theory concerning a certain principal connection on a principal bundle over spacetime and the covariant derivative operators induced by that connection on certain associated bundles.

Two remarks are in order before we return to Yang-Mills theory in the next section.  The first concerns what has been changed with this ``reformulation'' of general relativity.  In short, nothing.  In particular, no additional structure has been added to the theory.  Any manifold gives rise, in a canonical way, to an associated frame bundle.  Thus there is a straightforward sense in which a relativistic spacetime $(M,g_{ab})$ always comes equipped with a principal bundle over it; we just have little occasion to mention it in ordinary applications of relativity theory.  Similarly, relative to a fixed choice of isomorphism between $TM$ and an appropriate vector bundle associated to $LM$, the covariant derivative operators one regularly encounters in general relativity induce, or are induced by, a principal connection on $LM$.  One can even make the claim that nothing has been added precise, using a notion of relative amounts of structure recently discussed by \citet{Barrett}: the group of automorphisms of the frame bundle $LM$ that preserve the isomorphism between $TM$ and the bundle associated to $LM$ is canonically isomorphic to the diffeomorphism group of $M$.\footnote{See also \citet{Swanson+Halvorson} for a related discussion.}  This means, on Barrett's account, that the frame bundle (with fixed isomorphism) has precisely the same amount of structure as $M$.

The second remark brings us back to the promised moral of this section, which is that the frame bundle has a perfectly straightforward interpretation in the context of general relativity: as we noted when we defined it, it is the bundle of (ordered) bases of the tangent spaces at each point.  A section $\sigma:U\rightarrow LM$ of the frame bundle, then, is a \emph{frame field}, i.e., a smoothly varying assignment of a basis of the tangent space to each point of $U$.  Given one section, a change of section corresponds to a change of basis at each point.  A given section determines a (flat) derivative operator, namely, the one relative to which the frame field is constant.  The pullback of a principal connection $\omega^{\mathfrak{A}}{}_{\alpha}$ along a section $\sigma$ gives a representation of the connection relative to the corresponding frame, known as the connection coefficient associated with the frame; if the frame field is holonomic, i.e., if it corresponds to a choice of local coordinates on $U$, these connection coefficients are systematically related to the Christoffel symbols for $\nabla$, the derivative operator on $TM$ induced by the principal connection.

Given the status of frame fields and Christoffel symbols on modern geometrical approaches to relativity, the remarks above suggest that $LM$ and the structure defined on it are really auxiliary.  In other words, there is a reason that one need not mention a principal bundle or principal connection for most purposes in general relativity, which is that it is the induced covariant derivative operator on $M$ that matters.  It is this object that determines the trajectories of massive bodies and the dynamics of (tangent valued) matter fields.  Moreover, this derivative operator (and its associated curvature) may be fully and invariantly characterized without ever mentioning frames, gauges, connection coefficients, or anything of the sort.   The frame bundle merely provides an alternative---and for many purposes, less attractive---way of encoding information about this derivative operator.

It is in this sense that we have a deflationary interpretation of ``gauge transformations'', ``gauge invariance'', and related notions.  A choice of gauge in relativity theory---that is, a choice of local section of the frame bundle---is merely a choice of frame field relative to which one may represent certain geometrical facts, which one might just as well represent in other ways without ever mentioning gauge; the requirement of gauge invariance is just the requirement that the objects so-represented are not dependent on the choice of frame.  In other words, choosing a ``gauge'' is strongly analogous to choosing a coordinate system, and should be treated accordingly.

\section{Yang-Mills Theory Re-visited}\label{MainEvent}

In the previous section, we saw that one can think of general relativity as a theory of a principal connection on a principal bundle over spacetime.  When one does so, the principal bundle has a natural interpretation: it is the bundle of bases for the tangent spaces at each point.  One also sees a straightforward sense in which the frame bundle is auxiliary structure.  On the one hand, it is naturally definable in terms of the tangent spaces of the manifold, and on the other hand, it is still sections of the tangent bundle and other bundles constructed from it that we take to represent matter and its properties in spacetime.  One also sees that it is the induced covariant derivative operator on the tangent bundle, rather than the corresponding principal connection, that plays a direct role in the physics.  The principal bundle and principal connection, insofar as they play any role at all in general relativity, are best conceived as tools for coordinating structure that might also be defined directly on the tangent spaces.

Do similar considerations apply to Yang-Mills theory?  I claim they do.  First, note that although every manifold naturally gives rise to a frame bundle over that manifold, frame bundles are more general constructions.  In fact, \emph{any} vector bundle $V\rightarrow E \xrightarrow{\pi} M$ is associated with a corresponding frame bundle $GL(V)\rightarrow LE \rightarrow M$, which is a principal bundle over $M$, where $GL(V)$ is the group of automorphisms of $V$.\footnote{To be clear: $GL(V)$ is just the general linear group $GL(n,\mathbb{K})$, where $n$ is the (unspecified) dimension of $V$, and $\mathbb{K}$ is $\mathbb{R}$ or $\mathbb{C}$, depending on whether $V$ is real or complex. }  The typical fibers of these more general frame bundles consist in all bases of the fibers of $E$.  Moreover, the original vector bundle, $V\rightarrow E\xrightarrow{\pi} M$, is (isomorphic to) a vector bundle $V\rightarrow LE\times_{GL} V\xrightarrow{\pi_{\rho}} M$ associated to $LE$ with typical fiber $V$, determined by a faithful representation of $GL(V)$ on $V$.

It follows in particular that the vector bundles one encounters in the context of Yang-Mills theory, such as the bundle corresponding to complex scalar fields that we described in section \ref{Yang-Mills}, may also be understood as associated to frame bundles.  For instance, a bundle corresponding to a complex scalar field, $\mathbb{C}\rightarrow E\xrightarrow{\pi} M$, may be understood as a two dimensional (real) vector bundle associated to a $GL(2,\mathbb{R})$-principal bundle over $M$, $GL(2,\mathbb{R})\rightarrow LE\rightarrow M$.  Of course, when we originally introduced the vector bundle $E$, we introduced it as associated to a different principal bundle---namely, a $U(1)$-principal bundle, $U(1)\rightarrow EM\xrightarrow{\wp} M$.  These principal bundles bear a systematic relationship to one another: the $U(1)$ bundle is a principal sub-bundle of the frame bundle $L\mathbb{C}$, where the representation $\rho:U(1)\rightarrow GL(2,\mathbb{R})$ used to construct $E$ as an associated bundle of $EM$ determines the embedding of the structure group of $EM$ into the structure group of the frame bundle.  We may represent this relationship by defining the vertical principal bundle morphism $(\Psi_{\rho}, 1_M,\rho):(U(1)\rightarrow EM\xrightarrow{\wp} M)\rightarrow (GL(2,\mathbb{R})\rightarrow L\mathbb{C}\xrightarrow{\wp_L} M)$ that realizes the embedding of the $U(1)$ bundle in the frame bundle.  Note that nothing in this paragraph depended on the details of the $U(1)$ bundle or its $\mathbb{C}$ associated bundle: identical considerations apply to any principal bundle and associated vector bundles.\footnote{\label{faithful}In particular, it is the fact that $\rho$ is a \emph{faithful} representation of $U(1)$ that allows us to use it to produce a $U(1)$ subbundle of the frame bundle of the corresponding associated bundle.  This is not always possible: there exist representations of Lie groups that are not faithful,  and indeed, there exist Lie groups that do not admit faithful representations on any vector space.  Still, even in the most general case, a representation $\rho:G\rightarrow GL(V)$ (faithful or not) of a group $G$ on some vector space $V$ gives rise to \emph{some} subbundle of the frame bundle of the corresponding associated bundle, namely a $\rho[G]$ subbundle.  Moreover, in cases of interest in physics, one does have faithful representations, except where one uses trivial representations of a given Yang-Mills structure group as shorthand for the observation that a certain matter field is impervious to the interactions governed by that Yang-Mills force.}

Now fix some arbitrary principal bundle $G\rightarrow P\xrightarrow{\wp} M$ and an associated vector bundle $V\rightarrow P\times_{G} V\xrightarrow{\pi_{\rho}} M$, and assume $\rho:G\rightarrow GL(V)$ is faithful.  Several things follow from the observations of the previous paragraph.  The first is that a section $\sigma:U\rightarrow P$ of the principal bundle now has a straightforward interpretation: it may be understood as a frame field for $P\times_G V$.  This is because $P$ is naturally construed as a subbundle of $L(P\times_G V)$, the bundle of frames for $P\times_G V\xrightarrow{\pi_{\rho}} M$, with embedding $(\Psi_{\rho}, 1_M,\rho):(G\rightarrow P\xrightarrow{\wp} M)\rightarrow (GL(V)\rightarrow L(P\times_G V) \xrightarrow{\wp_L} M)$, which means that any section $\sigma$ of $P$ gives rise to a section $\Psi_{\rho}\circ \sigma:U\rightarrow L(P\times_G V)$ of the frame bundle.  Thus, a choice of ``gauge'' in the context of the $G$-bundle is simply a choice of frame field for $P\times_{G} V$; similarly, a change of gauge corresponds to a change in basis at each point.  A second observation uses the following fact: any connection $\tilde{\omega}^{\mathfrak{A}}{}_{\alpha}$ on $P$ extends uniquely to a connection $\omega^{\mathfrak{A}}{}_{\alpha}$ on $L(P\times_G V)$.\footnote{This holds generally for subbundles.  See \citet[Prop. 6.1]{Kobayashi+Nomizu}.}  Thus any connection on $P$ may be conceived as a connection on $L(P\times_G V)$, and so, in particular, given a connection $\tilde{\omega}^{\mathfrak{A}}{}_{\alpha}$ on $P$ and a section $\sigma$ of $P$, the pullback of the connection along $\sigma$ is precisely the corresponding connection on $L(P\times_G V)$ expressed in terms of connection coefficients in the frame $\Psi_{\rho}\circ \sigma$.  A third observation is that, though $P$ will in general have many associated bundles, the considerations above hold in every case.  Thus, one might see our $G$-principal bundle as a subbundle of several frame bundles, corresponding to different associated vector bundles; in each case, a principal connection on $P$ extends uniquely to a connection on the frame bundle, and a section of $P$ give rises to a frame field for all of the associated bundles.

These considerations suggest the following proposal: by understanding a principal bundle $P$ as a subbundle of the frame bundles of the vector bundles associated to it, we may interpret the principal bundles one encounters in Yang-Mills theory precisely as we argued one should interpret the frame bundle in general relativity, namely as bundles of local bases for various vector bundles.  But this proposal leads to an immediate question.  If we are going to interpret the principal bundles we encounter in Yang-Mills theory as bundles of frames for vector bundles, what is the significance of the fact that the structure groups that actually arise in the physics are not general linear groups?  In other words, if the proposal I have just made is reasonable, why do we restrict ourselves to \emph{subbundles} of frame bundles?

Here, too, general relativity provides a useful guide to interpretation.  In the previous section, we described general relativity in terms of a connection on $LM$, the bundle of all frames over a manifold $M$.  But a spacetime $(M,g_{ab})$ gives us the resources to identify some frames as distinguished.  These are the \emph{orthonormal} frames at each point, relative to the spacetime metric $g_{ab}$.  The bundle over spacetime of all orthonormal frames relative to $g_{ab}$, $O(1,3)\rightarrow FM\xrightarrow{\wp_F} M$, is a principal subbundle of $LM$, where the fibers are all isomorphic to the Lie group $O(1,3)$, the indefinite orthogonal group of degree $(1,3)$ or the \emph{Lorentz group}, which in turn is isomorphic to the group of invertible matrices that map one orthonormal basis to another.  (The converse is also true: given an $O(1,3)$ principal bundle over $M$, there exists a unique corresponding Lorentzian metric on $M$ relative to which the frames are orthonormal.)  Finally, a connection $\omega^{\mathfrak{A}}{}_{\alpha}$ on $LM$ induces a derivative operator $\nabla$ on $TM$ that is compatible with $g_{ab}$ in the sense that $\nabla_a g_{bc}=\mathbf{0}$ if and only if $\omega^{\mathfrak{A}}{}_{\alpha}$ is \emph{reducible} to a connection on $FM$, in the sense that there exists a connection $\tilde{\omega}^{\mathfrak{A}}{}_{\alpha}$ on $FM$ whose unique extension to $LM$ is $\omega^{\mathfrak{A}}{}_{\alpha}$.\footnote{Indeed, one can state the fundamental theorem of (pseudo)-Riemannian geometry in these terms: there exists a unique principal connection on any $O(1,3)$ principal bundle that induces a torsion-free derivative operator. See \citet[\S 6.2]{Bleecker}.}  Thus we see that connections on $FM$ may be thought of as connections on $LM$ that, in a precise sense, preserve the metric structure on $TM$ given by $g_{ab}$.  So in this case, at least, restricting attention to a subbundle of the frame bundle $LM$ is equivalent to endowing $TM$ with additional structure---namely, a metric---that is then preserved by any connections on that subbundle.

The same moral applies to the bundles we encounter in Yang-Mills theory.  In general, a subbundle $G\rightarrow P\xrightarrow{\wp} M$ of a frame bundle $GL(V)\rightarrow LE\xrightarrow{\pi_L} M$ corresponds to a restriction of the general linear group on $V$ to those automorphisms of $V$ that also preserve some additional structure---which, as in the case of a Lorentzian metric, turns out to be equivalent to defining such a structure on $V$.\footnote{To be clear, a given structure on $V$ always determines a subbundle of $LE$; the converse is true for each fiber of the vector bundle, but in some cases a further integrability condition is needed to define the corresponding structure on the entire vector bundle.}  For instance, as we have seen, $O(p,q)$ subbundles of the frame bundle of a vector bundle with $(p+q)$ dimensional fibers arise from, and determine, a non-degenerate $(p,q)$ signature metric.  An $SL(n,\mathbb{R})$ subbundle of a $GL(n,\mathbb{R})$ bundle, where $SL(n,\mathbb{R})$ is the real special linear group of degree $n$, which is the subgroup of $GL(n,\mathbb{R})$ corresponding to invertible $n\times n$ matrices of determinant 1, determines a volume form on a vector bundle with $n$ dimensional fibers.  And so on.  In all such cases, reducible connections preserve this additional structure.

Although Yang-Mills theory is, in principle, defined for principal bundles of any structure group, the examples one encounters in the Standard Model and various Grand Unified Theories are of special interest, so it is worth pausing to mention how these structure groups correspond to additional structure on their associated bundles.\footnote{What is the physical significance of this additional structure?  It is hard to say, for the same reasons that I expressed dissatisfaction in the introduction and fn. \ref{dis2}.  Ultimately, the structure is a relic of the fact that we are interpreting fields associated with quantum states---where a Hermitian inner product is natural---as classical matter.}  A $U(n)$ subbundle of a $GL(2n,\mathbb{R})$ bundle determines an $n$ dimensional complex vector space structure and a Hermitian inner product on a $2n$ dimensional (real) vector space.\footnote{Of course, one can consider $U(n)$ subbundles of the frame bundles of vector bundles with higher dimensional fibers; in such cases, the $U(n)$ subbundle puts a complex vector space structure and a Hermitian inner product on a $2n$ dimensional subspace of the fibers.  Similar considerations apply in the other cases mentioned.}  Meanwhile, an $SU(n)$ subbundle of a $GL(2n,\mathbb{R})$ bundle determines an $n$ dimensional complex vector space structure, a Hermitian inner product, and an orientation on a $2n$ dimensional real vector space.  Finally, an $SO(n)$ subbundle of a $GL(n,\mathbb{R})$ bundle determines a positive definitive metric and an orientation on an $n$ dimensional real vector space.

Returning now to the proposal above, we see that the fact that the principal bundles we most often encounter in Yang-Mills theory do not have the general linear group of any vector space as their structure group does not block interpreting them as (sub)bundles of frames for a given collection of vector spaces.  In fact, we see now that they are bundles of frames that are appropriately compatible with additional structure on these vector spaces.  Similarly, connections on these bundles are connections on the frame bundle that preserve this additional structure, in the sense that the Levi-Civita derivative operator preserves the metric in general relativity.  We are thus led to a picture on which we represent matter by sections of certain vector bundles (with additional structure), and the principal bundles of Yang-Mills theory represent various possible bases for those vector bundles.

Following the line of reasoning offered at the end of the previous section, then, it is tempting to conclude that the principal bundles of Yang-Mills theory are merely auxiliary structure.\footnote{This perspective, which is very congenial to the one offered here, is also reflected in \citet{Palais}; Geroch (private correspondence) appears to take the same line.  I should emphasize, though, that on a purely mathematical level, the facts on the ground support a kind of equanimity regarding ``which comes first'', the principal bundle or the vector bundles.  My claim here is that, with regard to the \emph{physics}, the formalism is much more naturally understood if one takes the vector bundles to be primary, since they play a more direct role in representing matter.}  Principal bundles do not represent states of matter, nor do they represent space or time; instead, they represent bases for certain vector spaces.  From this point of view the standard terminology is misleadingly backward: it is the vector bundles that matter, and for some purposes, one might also introduce associated principal bundles.  The real physical significance should be attached to (sections of) the associated vector bundles and the covariant derivative operators acting on them.  As in general relativity, all of this structure may be characterized fully and invariantly without mentioning frames, gauges, or connection coefficients.  And again, also as in general relativity, these considerations lead to a deflationary attitude towards notions related to ``gauge'': a choice of gauge is just a choice of frame field relative to which some geometrically invariant objects---the derivative operators on a vector bundle, say---may be represented, analogously to how geometrical objects may be represented in local coordinates.

But this does not quite mean one can forget about the principal bundles altogether, as it seems one can forget about the frame bundle in general relativity.  The reason is that, in general relativity, all of the bundles associated to the frame bundle---that is, the tangent bundle, the cotangent bundle, and bundles of tensors acting on them---can be defined directly in terms of tensor products of the tangent bundle and cotangent bundle, which in turn can be defined directly in terms of tangent vectors and linear functionals.  This method of constructing the bundles allows one to take a covariant derivative operator on the tangent bundle and immediately induce a corresponding derivative operator on all of the other vector bundles one cares about, without ever mentioning that they all correspond to a single principal connection on the frame bundle.

In Yang-Mills theory, however, the situation is a bit different.  To see why, observe that charged scalar fields, electron fields, muon fields, tau particle fields, quark fields of various sorts, etc., all interact electromagnetically---and more importantly, in more traditional language, they all respond to the \emph{same} electromagnetic fields.  This means that the covariant derivative operators on the vector bundles in which these fields are valued must be, in some appropriate sense, the same, since all of these (different) vector bundles must have the same parallel transport and curvature properties.  But unlike in the case of the tensor bundles just described, these vector bundles bear no direct relationship to one another, except that they have, in some sense, the same derivative operator.

This is where the Yang-Mills principal bundles become important: that electrons, muons, etc. are all represented by vector bundles associated to the \emph{same} $U(1)$-principal bundle, and have covariant derivative operators induced by the same principal connection, provides the precise sense in which these different kinds of particles all respond to the same electromagnetic influences.  In effect, the principal bundles in Yang-Mills theory coordinate frames across different vector bundles.  And this coordination of frames allows one to make precise the senses in which (1) vectors in different vector spaces might be constant by the same standard of constancy and (2) different vector bundles might have the same (dynamical) curvature.  Thus, even if one takes principal bundles to be in some sense auxiliary or secondary as far as the physics is concerned, they still play an important role in Yang-Mills theory, in a way that the frame bundle need not in general relativity.\footnote{One might worry that the role I have just ascribed to the principal bundles---of coordinating physically significant data concerning parallel transport and curvature across different kinds of charged matter---is robust enough that it is misleading to call it ``auxiliary'', since being ``auxiliary'' may suggest that the structure is unnecessary or eliminable.  In any case, I hope I have been clear enough above about what I take the roles of the various bundles to be---vector bundles represent possible local states of matter; principal bundles coordinate between these vector bundles---that the sense of ``auxiliary'' I have in mind is clear.  It is the sense in which a coach is auxiliary to the players on the field.}

%The picture that emerges is one on which some matter has properties that are best conceived as ``vectorial''---examples might include velocity, electromagnetic charge, spin, isospin, color charge, etc.  We represent this matter by various vector fields on spacetime, such as the two-component field representing a charged ``scalar'' field, or the six component field representing (classical) quark fields.  A given distribution of matter may be represented by several such fields, as in the case where one represents a charged fluid by its mass density and its charge density, among other properties (such as its 4-velocity density).  In some cases, one might freely think of these vector fields as valued in the tangent bundle, but in general this is not possible, for several reasons: (1) there is no reason to expect that generic properties will be valued in vector spaces with the same dimension as spacetime; (2) the vector spaces in which these properties are valued may have more structure than the tangent space, such as a complex vector space structure; (3) conversely, the vectors spaces in which these properties are valued may have less structure than the tangent space, since, for instance, there may be no natural partition of such properties into spacelike and causal; and (4) the vectors associated with properties need not have any interpretation in terms of ``directions'' in spacetime, i.e., in terms of tangents to curves in spacetime.

\section{Disanalogies: Real and Not so Much}\label{Disanalogies}

The last section ended with the observation that, even if one interprets the principal bundle formalism in Yang-Mills theory as one would in general relativity, there is still an important difference regarding how one ultimately views the  principal bundles and principal connections that one encounters in each case.  In general relativity, the principal bundles are largely dispensable, whereas in Yang-Mills theory, even if one takes the view that the principal bundles are auxiliary structure, that auxiliary structure is needed, since it is what allows us to coordinate curvature across different vector bundles that are not otherwise related.  This disanalogy between Yang-Mills theory and general relativity does not threaten the main arguments of the previous section, regarding how one should best understand the geometry of Yang-Mills theory on a ``fiber bundle interpretation''.  But it does underscore the fact that, analogies notwithstanding, Yang-Mills theory and general relativity are also importantly \emph{disanalogous}.  We now turn our attention to some other disanalogies, which some authors---notably, \citet{Trautman}, \citet{Anandan}, and \citet{HealeyOld, Healey}---have argued block the general style of argument presented here, of using one's interpretation of general relativity as the basis for an interpretation of Yang-Mills theory.  I will argue that the significance of these disanalogies has been overstated, at least for the proposal made in the present paper.

A first disanalogy, which is both crucial to the physics of both theories and irrelevant to the present discussion, is that Einstein's equation is not an instance of the Yang-Mills equation.  Both equations may be understood to relate the curvature of a principal bundle to properties of matter, but the relationships are simply not the same.  But the fact that these are distinct theories, or that one is not a special case of the other, does not undermine the claim that they are analogous in certain important ways---particularly because the analogies on which the arguments of the previous section depend concern general properties of principal bundles, principal connections, and associated bundles that are independent of the dynamics of how a connection or derivative operator is related to the matter in spacetime.\footnote{Note that the reason that the dynamics of these theories are different---or even can be different, given the uniqueness results discussed in \citet[\S 10.2]{Bleecker}, \citet[pp. 80-2]{Palais}, or \citet{Feintzeig+Weatherall}---is intimately related to the other disanalogies discussed in the present section, and so in that sense the differences in the dynamics may be relevant to the discussion here.  But the mere fact that they \emph{are} different is not relevant.}

A second disanalogy that is sometimes noted is actually illusory: namely, that in general relativity, we have a (dynamical) metric on the tangent space, rather than just a derivative operator, whereas in Yang-Mills theory, we have only a dynamical connection. There are two ways in which this apparent disanalogy is misleading.  The first way is that, as can be seen from the discussion of the Yang-Mills equation in section \ref{Yang-Mills} and the Hodge star operator in \ref{AppHodgeStar}, the spacetime metric plays an essential role in defining the dynamics of Yang-Mills theory.  So it would be misleading to say that the spacetime metric appears only in general relativity.  Both theories require a spacetime metric.   The second way in which this apparent disanalogy is misleading is that, although the spacetime metric defines a preferred inner product on fibers of the tangent bundle, but not on fibers of other bundles, all of the vector bundles that one encounters in practice in Yang-Mills theory are associated to principal bundles with unitary or orthogonal structure groups, which, as we saw in section \ref{MainEvent}, implies that there \emph{is} an inner product on the fibers of these associated bundles.  Indeed, standard methods of Lagrangian field theory \citep[see][Ch. 4]{Bleecker} take such an inner product for granted, suggesting that that is also a sense in which Yang-Mills theory as standardly understood should be construed to require some inner product on the vector bundles representing states of matter.

A third disanalogy is real, and it is important.  There are several ways of putting this point, but they all boil down to the following fact: in general relativity, the associated bundles of interest are the \emph{tangent} bundle and related bundles constructed from it; in Yang-Mills theory, the vector bundles are not tangent to the manifold, in the sense that elements of the bundles do not represent tangent vectors to curves at a point.  This matters because the relationship between a manifold $M$ and its tangent bundle $TM$ endows $TM$ with structure that one does not generally have on a vector bundle, related to the fact that the elements of $TM$ may be conceived as tangent vectors to smooth curves in $M$, or equivalently, as derivations on smooth functions $\varphi:M\rightarrow \mathbb{R}$ on $M$.  For instance, a section of the tangent bundle may be associated with a family of integral curves on a manifold, which in turn may be used to determine a one-parameter family of local diffeomorphisms from (open sets of) $M$ to (open sets of) $M$.  Likewise, given a smooth map $f:M\rightarrow N$, one can immediately define pushforward and pullback maps between (fibers of) $TM$ and $TN$ by making use of the relationship between tangent vectors and derivations on smooth functions. Given a generic vector bundle $E\xrightarrow{\pi} M$ over $M$, meanwhile, a map $f:M\rightarrow N$ does not define a pushforward of any sort---even if one has a vector bundle defined over $N$.\footnote{If one has a vector bundle $E'\rightarrow N$ over $N$, a smooth map $f:M\rightarrow N$ may be used to define a \emph{pullback bundle}, $f^*(E')\rightarrow M$, which is a bundle whose fiber at each point $p\in M$ is the fiber at $f(p)$ of $E'$.  But this construction defines a \emph{new} bundle; it does not generate a map between a fixed bundle over $M$ and a bundle over $N$.}

The structure described in the last paragraph, relating sections of the tangent bundle and smooth maps, allows one to define the Lie derivative of one section of the tangent bundle relative to another:
\[
\mathcal{L}_{\eta}\xi^a = \lim_{t\rightarrow0}\frac{1}{t}\left(((\Gamma_t)^*(\xi^a))_{|p} - (\xi^a)_{|p}\right),
\]
where $\xi^a$ and $\eta^a$ are smooth local sections of the tangent bundle with overlapping domains, and $\{\Gamma_t\}$ is a local one-parameter family of diffeomorphisms generated by $\eta^a$.  The Lie derivative in turn allows one to define the \emph{torsion} of a covariant derivative operator $\nabla$ on $TM$, $T^a{}_{bc}$, which is the field on $M$ (or, the section of the appropriate bundle of tensors) with the property that, given any two sections of $TM$, $\xi^a$ and $\eta^a$,
\[
T^a{}_{bc}\eta^b\xi^c = \eta^n\nabla_n\xi^a - \xi^n\nabla_n\eta^a - \mathcal{L}_{\eta}\xi^a.
\]
The torsion is a measure of the degree to which parallel transport along a curve, relative to $\nabla$, implements a rigid rotation of the tangent space, relative to the standard set by the Lie derivative.  Since Lie derivatives do not exist for sections of arbitrary vector bundles, one cannot define a torsion tensor for derivative operators on arbitrary vector bundles.  Thus, there is a sense in which derivative operators on the tangent bundle have structure that generic derivative operators do not have.

The existence of a torsion tensor captures part of how tangent bundles differ from generic vector bundles.  The torsion tensor has been discussed in connection with the differences between general relativity and Yang-Mills theory by, for instance, \citet{Trautman} in the context of a discussion of the possible field equations for generalizations of general relativity inspired by the principal bundle formalism.  In that context, it is salient that there is an additional object---the torsion tensor---that might be constrained by the dynamics of one's theory.  Indeed, there is a sense in which general relativity already makes use of the torsion tensor, albeit in a backhanded way: it is by requiring that the derivative operator be torsion-free that one secures a \emph{unique} derivative operator compatible with a given spacetime metric.\footnote{Compare this with the situation in the Yang-Mills theories we have described, where one generally has a non-degenerate inner product on the fibers of one's bundle, but where there are generally many connections compatible with this inner product.  The reason is that there is no analogue of torsion for a connection on a generic vector bundle, and so there is no way to single out an analogue to the Levi-Civita derivative operator.}  But one can certainly entertain the possibility of allowing torsion to play a more substantive role.

Thus, there is a sense in which the existence of a torsion tensor is an important disanalogy between Yang-Mills theory and general relativity or other theories of gravitation, insofar as it provides additional degrees of freedom for constructing field equations.  But it is not clear that this particular disanalogy affects the arguments of the previous section.  The existence of a torsion tensor allows one to develop a richer theory of derivative operators in the context of the tangent bundle of a manifold, but it does not follow that torsion changes how one should interpret derivative operators in the first place, nor does it have any bearing on the relationship between vector bundles, frame bundles, and principal bundles more generally, which was what the interpretation above relied on.

That said, there is another object, closely related to the torsion tensor, that, it has been argued, does bear on the relationships on which the arguments of the previous section depend. This object is known as a \emph{solder form}.  The existence of a solder form on the frame bundle associated with general relativity has led some commentators---including \citet{HealeyOld, Healey} and \citet{Anandan}, on whom Healey relies---to conclude that one simply cannot interpret the principal bundle formalism in Yang-Mills theory in the way one would in general relativity.  Some subtlety is called for in this discussion, however.  There is a strong sense in which the solder form itself is a red herring, at least on the interpretation of the principal bundle formalism defended here.  But the difference between general relativity (or gravitational theories more generally) and Yang-Mills theory that Anandan ultimately (and misleadingly) takes the solder form to be emblematic of is real and significant.  Nonetheless, I will argue, once one is clear about what structure Anandan's argument relies on, it will follow that the interpretation described in the present paper is unaffected.

To define the solder form, consider an $n$ dimensional manifold $M$ and the frame bundle $GL(n,\mathbb{R})\rightarrow LM\xrightarrow{\wp_L} M$ over $M$.  Recall that the tangent bundle $V\rightarrow TM\xrightarrow{\pi_T} M$ over $M$ is isomorphic to a vector bundle $V\rightarrow LM\times_{GL} V\xrightarrow{\pi_{\rho}} M$ associated to $LM$.  It was this isomorphism between $TM$ and $LM\times_{GL} V$ that supported our taking the tangent bundle to be a vector bundle associated to the frame bundle; it was also what allowed us to associate structure on the frame bundle with structure directly on the tangent bundle, as when we considered the Levi-Civita derivative operator as a connection on the principal bundle---in effect, we used this isomorphism to assign a torsion tensor to a principal connection on $LM$, which is not possible for general principal connections.  In general, however, this isomorphism is not unique, and so some choice must be made, relative to which the constructions just described proceed.  Usually the isomorphism is taken to be fixed once and for all. A \emph{solder form} is some such choice of vector bundle isomorphism between $TM$ and $LM\times_{GL}V$.

A solder form may be represented as a smooth mixed index tensor field on $M$, $\vartheta^A{}_{a}$, with inverse $\vartheta^a{}_A$, whose action at each point $p$ defines a linear bijection between $T_pM$ and $(\pi_{\rho})^{-1}[p]$, the fiber of $LM\times_{GL} V$ at $p$; alternatively (and equivalently), it may be represented by a mixed index tensor field on $LM$, $\bar{\vartheta}^A{}_a$, with inverse $\bar{\vartheta}^a{}_A$, whose action at each point $u\in LM$ defines a bijection between $T_{\wp_L(u)}M$ and $V$, the (fixed) vector space relative to which $LM\times_{GL} V$ is defined.  Here $\bar{\vartheta}^A{}_a$ is required to be equivariant with respect to the $GL(n,\mathbb{R})$ action on $LM$ in the sense that $(\bar{\vartheta}^A{}_a)_{|ug}=(\rho(g^{-1}))^A{}_B(\bar{\vartheta}^B{}_a)_{|u}$,\footnote{Here $(\rho(g^{-1}))^A{}_B$ is the tensor corresponding to the action of $g^{-1}$ on $V$ in the representation $\rho$.} for any $g\in GL(n,\mathbb{R})$. It is in this latter form that the solder form is most often encountered, and so we will focus on this form in what follows.

The solder form is often treated as ``canonical'' or ``tautological'', in the sense that a particular solder form is naturally definable using the structure of the frame bundle---indeed, various sources, such as \citet{Bleecker} and \citet{Kolar+etal} simply call it the canonical one form.  The idea is that, from the construction of $LM$, each point $u\in LM$ corresponds to a particular basis $u=\{\overset{1}{u}{}^a,\ldots,\overset{n}{u}{}^a\}$ for $T_{\wp_L(u)}M$.  Thus, given some fixed basis $v=\{\overset{1}{v}{}^A,\ldots,\overset{n}{v}{}^A\}$ for $V$, one gets a solder form on $LM$ by defining, for each $u\in LM$, $(\bar{\vartheta}^A{}_a)_{|u}\overset{i}{u}{}^a=\overset{i}{v}{}^A$---that is, one defines a solder form by assigning to each point $u\in LM$ the linear bijection that takes the basis $u$ of the tangent space at $\wp_L(u)$ to $v$.  This certainly does define a solder form: it is a smoothly varying linear bijection at each point of $LM$ and it is equivariant in the required sense, so it generates a vector bundle isomorphism between $TM$ and $LM\times_{GL} V$.  But it is only ``canonical'' relative to a number of prior choices, including a choice of basis for $V$ and an identification of points of $LM$ (which, after all, is just some manifold) with ordered $n$ tuples of vectors in $TM$.  The solder form itself is the object that fixes how these choices relate to one another, and so it is canonical just insofar as one has certain such choices in mind.  Of course, the construction of the frame bundle suggests certain choices, and the ``canonical'' solder form is the one that reflects those ``natural'' choices (modulo a choice of basis on $V$).%\footnote{More generally, one might think of the maps from $LM$ to $V$ used to define an associated bundle as sections of the trivial bundle $E\cong LM\times V\xrightarrow{\pr_1} LM$, relative to a fixed global trivialization.  Then the solder form is canonical only relative to this particular choice of global trivialization and a basis on $V$.  This additional freedom matters when one considers how to think about solder forms on isomorphic principal bundles.}

Given the discussion in the last paragraph, it may be unsurprising that the solder form has the property that it is not invariant under any non-trivial vertical principal bundle automorphisms,\footnote{The reason this should be unsurprising is that vertical principal bundle automorphisms will generally not preserve the ``choices'' described above.} in the sense that given any such automorphism $(\Psi,1_M,1_{GL}):(LM\xrightarrow{\wp_L} M)\rightarrow (LM\xrightarrow{\wp_L} M)$, $(\bar{\vartheta}^A{}_a)_{|u}\neq (\bar{\vartheta}^A{}_a)_{|\Psi(u)}$.\footnote{\label{soldertransform} Note that this expression makes sense, since both sides of the inequality are evaluated at points of the same fiber, and at every point of that fiber, $\bar{\vartheta}^A{}_a$ is a map between the same two vector spaces, viz. $T_{\wp_L(u)}M$ and $V$.  To see that the inequality holds, note that for any $u\in LM$, there is some $g\in GL(n,\mathbb{R})$ such that $\Psi(u)=ug$, where $g=e$ iff $\Psi=1_{LM}$.  And since $\bar{\vartheta}^A{}_a$ is equivariant, $(\bar{\vartheta}^A{}_a)_{|\Psi(u)}=(\bar{\vartheta}^A{}_a)_{|ug}= (\rho(g^{-1}))^A{}_B(\bar{\vartheta}^B{}_a)_{|u}\neq(\bar{\vartheta}^A{}_a)_{|u}$.}  Indeed, we already made implicit use of this fact in section \ref{GR}, when we argued that defining a frame bundle added no structure to a manifold.\footnote{It is interesting to note that without a choice of solder form on $LM$, the group of principal bundle automorphisms is naturally a supergroup of the group of diffeomorphisms of its base space.  So on the \citet{Barrett} account, a frame bundle without a solder form has \emph{less} structure than its base space.}  Nonetheless, this property is striking if one thinks of the solder form as canonical, since it would appear to mean that a frame bundle has fewer automorphisms than one would expect of an ordinary principal bundle with the same structure group. Anandan puts the point dramatically: ``the ... gauge symmetry of the gravitational field is broken by the existence of the solder form'' \citep[p. 10]{Anandan}.  For this reason, Anandan and Healey argue, gravitational theories based on the frame bundle of a manifold are fundamentally different from Yang-Mills theory.\footnote{There is a tension in Anandan's view, here.  He takes the solder form to be canonical, in the sense that it is a structure that both arises naturally on $LM$ and ``breaks'' the symmetry; but he also argues that the considerations he presents suggest that the solder form should be a fundamental \emph{dynamical} variable in theories of gravitation, implying that different configurations of matter would lead to different solder forms.  One cannot have it both ways.}

It is hard to know what to make of such claims.  Perhaps the most important reason to be puzzled is that although it is true that a solder form is not preserved by vertical principal bundle automorphisms, neither, in general, is a principal connection.  And so, insofar as a model of Yang-Mills theory requires one to specify a connection, the ``gauge symmetry is broken'' in generic models of every Yang-Mills theory in precisely the same sense that Anandan claims it is broken in theories based on $LM$.\footnote{There is an analogy here to general relativity in a different guise.  In that context, one often hears that general relativity exhibits ``diffeomorphism invariance''.  But it is not the case that, given a relativistic spacetime $(M,g_{ab})$ and a diffeomorphism $f:M\rightarrow M$, $f^*(g_{ab})=g_{ab}$.  So in this sense---which is the same as the sense in which the solder form ``breaks'' ``gauge symmetry''---the metric ``breaks'' diffeomorphism invariance.  What \emph{is} the case is that if $(M,g_{ab})$ is a spacetime, then $(M,f^*(g_{ab}))$ is also a spacetime, and it is ``equivalent'' in the sense that $(M,g_{ab})$ and $(M,f^*(g_{ab}))$ may be used to represent the same physical situations.  But this, mutatis mutandis, is precisely what happens with the solder form.  See \citet{WeatherallHoleArg} for a discussion of related issues and how they have led to spurious arguments in the relativity literature.}  On the other hand, given a principal connection $\omega^{\mathfrak{A}}{}_{\alpha}$ on a principal bundle $G\rightarrow P\xrightarrow{\wp} M$ and a vertical principal bundle automorphism $(\Xi,1_M,1_G):(P\xrightarrow{\wp} M)\rightarrow(P\xrightarrow{\wp} M)$, $\Xi^*(\omega^{\mathfrak{A}}{}_{\alpha})$ is also a principal connection on $P$ that is equivalent to $\omega^{\mathfrak{A}}{}_{\alpha}$ in the sense that it may be used to represent all of the same physical situations.  But solder forms behave in just the same way: given a vertical principal bundle automorphism $(\Psi,1_M,1_{GL}):(LM\xrightarrow{\wp_L} M)\rightarrow (LM\xrightarrow{\wp_L} M)$, if $\bar{\vartheta}^A{}_a$ is a solder form, then $\Psi^*(\bar{\vartheta}^A{}_a)$ is also a solder form, where $\Psi^*(\bar{\vartheta}^A{}_a)$ is the mixed index tensor field on $LM$ whose action at $x\in LM$ is the same as the action of $\bar{\vartheta}^A{}_a$ at $\Psi(x)$.

All of that is to say that it is not perfectly clear why the existence of a solder form on $LM$, or even the specification of a particular solder form, leads to some special problem in drawing the analogies on which the arguments of section \ref{MainEvent} rely.  But in fact, there is a stronger reason to think that the solder form on $LM$ is irrelevant for present purposes, which is that there is a sense in which one has solder forms on \emph{all} frame bundles, not just frame bundles associated to the tangent bundle.  Suppose one is given a vector bundle $V\rightarrow E\xrightarrow{\pi} M$ and one constructs a frame bundle for that vector bundle, $GL(V)\rightarrow LE\xrightarrow{\wp_L} M$.  We know that the associated vector bundle $LM\times_{GL} V\xrightarrow{\pi_{\rho}} M$ is isomorphic (as vector bundles) to $E$.  But this isomorphism is not unique.  And so, for precisely the reasons that have been discussed, one needs to specify a particular isomorphism---that is, a solder form.  This isomorphism, too, may be represented by a suitably equivariant mixed index tensor on $LE$, $\theta^{\bar{A}}{}_{A}$,\footnote{Here the $\bar{}$ over the raised $A$ index indicates that this index is valued in a different vector space than the lowered index.  (Neither space is a tangent space to any manifold.)} which at each point $u\in LE$ yields a linear bijection between the fiber of $E$ at $\wp(u)$ and the fiber of $LM\times_{GL} V$ at $\wp(u)$.  The field $\theta^{\bar{A}}{}_{A}$ is what identifies a point $u\in LE$ with a particular basis for the fiber of $E$ at $\wp(u)$, in precisely the same way that $\bar{\vartheta}^A{}_a$ does for $LM$.  Moreover, there is a ``canonical'' choice of solder form on $LE$ in precisely the same sense as $LM$.  And like $\bar{\vartheta}^A{}_a$, $\theta^{\bar{A}}{}_{A}$ is not invariant under non-trivial vertical principal bundle automorphisms.\footnote{One can extend this argument to \emph{any} principal bundle with an associated bundle.  Given a principal bundle $G\rightarrow P\xrightarrow{\wp} M$, a vector space $V$, and a representation $\rho:G\rightarrow V$, one can always define an equivariant mixed index tensor $\delta^{\bar{A}}{}_B$ on $P$ whose action at each $x\in P$ is to map vectors in the fiber of $P\times_{G} V$ at $\wp(x)$ to the vector in $V$ corresponding to it.  This, too, is a kind of solder form: it specifies how $P$ relates to its associated bundle. But it is not invariant under vertical principal bundle automorphisms of $P$!}  So on the proposed interpretation of Yang-Mills theory given here, the existence or salience of a solder form on a principal bundle is not a disanalogy at all.

This does not mean that Anandan and Healey do not draw attention to a real disanalogy.  The real disanalogy is not that there \emph{is} a solder form on $LM$; rather, it is that $LM$ is soldered to the \emph{tangent bundle}.  (Of course, this is just a restatement of the point made above, that ultimately the difference between general relativity and Yang-Mills theory is that the vector bundles one encounters in general relativity are tangent bundles.)  We have already seen some reasons why this difference matters.  Another reason that is sometimes cited concerns the properties of parallel transport in a vector bundle relative to a covariant derivative operator, as drawn out by an example due to Anandan.  It is this example that Anandan and Healey take to be dispositive, and so we will end this section with a detailed discussion of Anandan's example.

To begin, suppose one has a vector bundle $E\xrightarrow{\pi} M$ and a covariant derivative operator $\nabla$ on $E$.  Then given a point $p\in M$, a vector $(\xi^A)_{|p}\in\pi^{-1}[p]$, and a curve $\gamma:I\rightarrow M$ with tangent field $\eta^a$ such that $p\in \gamma[I]$, one can consider the parallel transport of $\xi^A$ along $\gamma$.  For simplicity, suppose $[0,1]\subseteq I$ and $\gamma(0)=p$.  The perverse might be inclined to ask the following question: is $(\xi^A)_{|\gamma(0)}$ the same as $(\xi^A)_{|\gamma(1)}$?  But this question is ill formed, because the vectors $(\xi^A)_{|\gamma(0)}$ and $(\xi^A)_{|\gamma(1)}$ live in different vector spaces.  We have no way to compare them, except by parallel transporting both vectors along some curve to the same point.  In other words, the only way to determine whether $(\xi^A)_{|p}$ \emph{changes} when parallel transported along a curve is to consider parallel transport around closed curves.  Of course, this is not to say that there is any ambiguity concerning the parallel transport of $(\xi^A)_{|p}$ along any given curve, whether the curve is open or closed.  Rather, the point is that certain ways of talking about parallel transport along open curves are ambiguous.

The observations of the previous paragraph hold generically.  But, as Anandan observed, the situation is slightly different in the special case of parallel transport in the tangent bundle---at least when one has a metric $g_{ab}$.\footnote{In the general case, a metric is no help.} One still cannot say unambiguously whether $(\xi^a)_{|\gamma(0)}$ is ``the same'' as $(\xi^a)_{|\gamma(1)}$.  (Here we have changed indices to reflect the fact that $\xi^a$ is a tangent vector.)  But one can ask whether $\xi^a$ has rotated \emph{relative to $\eta^a$}, the tangent to the curve along which one has parallel transported $(\xi^a)_{|p}$---i.e., one can ask whether $(\xi^a\eta^bg_{ab})_{|\gamma(0)} = (\xi^a\eta^bg_{ab})_{|\gamma(1)}$.  This is because in the case of the tangent bundle, the vector tangent to the curve along which one is parallel transporting lives in the same vector space at each point of the curve as the vector being parallel transported.  And so in this case, there is a sense in which one can say a bit more about how vectors change when parallel transported along open curves, relative to a standard of constancy given by the tangent to the curve along which one has parallel transported.

The relevant question to ask, here, is whether the observations of the previous paragraph lead to some interesting notion of ``holonomy'', or more roughly ``change induced by parallel transport,'' for open curves. Anandan seems to suggest it does, by focusing attention on parallel transport around closed radial curves $\gamma:[0,1]\rightarrow C$ on a (two dimensional) cone $C$, with the standard metric $g_{ab}$ and derivative operator $\nabla$.\footnote{This case is of interest because it is often cited as an analogue to the Aharanov-Bohm effect, wherein a quantum particle propagating around a solenoid exhibits a distinctive interference pattern even though the electromagnetic field vanishes in the region in which the particle propagates. The reason the cone is analogous is that, with the standard metric and derivative operator, the cone is everywhere flat, but because it is not simply connected, parallel transport is globally path-dependent. Likewise, in the Aharanov-Bohm effect, one has non-trivial phase shifts corresponding to non-trivial ``holonomies'' of the principal connection on the principal bundle associated with electromagnetism, even though the electromagnetic field (i.e., the curvature of the principal connection) vanishes everywhere.  See \citet[Ch.2]{Healey} for an extensive discussion of the Aharanov-Bohm effect.}  In this case, where (1) the tangent field to $\gamma$, $\eta^a$, is such that $(\eta^a)_{|\gamma(0)}=(\eta^a)_{|\gamma(1)}$ and (2) a vector $\xi^a$, parallel transported from $\gamma(0)$ to $\gamma(1)$ along $\gamma$, is rotated relative to $\eta^a$ by an angle $\theta$, one can make precise a sense in which $\xi^a$ rotates continuously along $\gamma$, by observing that for any $t\in [0,1]$, the parallel transport of $(\xi^a)_{|\gamma(0)}$ to $\gamma(t)$ will have rotated by $t\theta$ relative to $\eta^a$.

Compelling as this particular example might be, its interest depends on certain very special features of the cone.  For instance, the fact that the cone has some curves that are ``privileged'' or ``constant'' (insofar as they are of constant distance from the apex at each point) but which are not geodesics plays an essential role, since if $\gamma$ \emph{were} a geodesic, then there would be no relative rotation between $\xi^a$ and $\eta^a$ along the curve,\footnote{This is because in such a case, $\eta^n\nabla_n(g_{ab}\xi^a\eta^b)=0$, meaning that the angle between $\xi^a$ and $\eta^a$ is constant everywhere on $\gamma$.  This is one sense in which the metric is preserved by $\nabla$.} and if $\gamma$ were in no sense constant, it would be unclear why $\eta^a$ should count as a salient standard of constancy.  Likewise, the fact that the manifold is two dimensional, so that a single angle is sufficient to uniquely characterize any vector relative to $\eta^a$, is what allows one to fully characterize how $\xi^a$ changes along $\gamma$ by appeal only to its properties relative to $\eta^a$; in the general case, $\xi^a$ could rotate wildly in a plane orthogonal to $\eta^a$, and the angle relative to $\eta^a$ would not record the change.  Thus, it is far from clear that in the general case, the change in angle of a parallel transported vector relative to the (arbitrary) curve along which is it parallel transported measures any quantity of special interest, or could stand in as a notion of ``holonomy'' along open curves.\footnote{For similar reasons, it is not clear that this sense of holonomy along open curves is sufficient to support Healey's conclusion that general relativity is ``separable,'' in the sense that parallel transport depends only on local properties, in a way that Yang-Mills theory is not.  (A similar point is made by \citet{Myrvold}.) But I will defer further discussion of the relationship between the present views and Healey's ``holonomy interpretation'' of Yang-Mills theory to future work.  (See, in particular, \citet{Rosenstock+Weatherall1, Rosenstock+Weatherall2}.)}  And in any case, even if in some special cases there is a difference of this sort, it should be clear that it is entirely orthogonal to the analogies between general relativity and Yang-Mills theory that motivated the interpretation presented in section \ref{MainEvent}.

\section{Conclusion}

Stepping back from the details, one might describe general relativity as follows. It is a theory according to which spacetime is curved, where the curvature depends on the distribution of energy and momentum in space and time, and gravitational effects are a manifestation of this spacetime curvature.  At the same level of description, the picture of Yang-Mills theory that one ends up with, on the view described here, is as follows.  The matter in the universe, consisting of electrons, quarks, neutrinos, and so on, seems to have degrees of freedom at each point that are well-represented by elements of a vector space.  Taken together, the spaces of local degrees of freedom form a vector bundle over spacetime, sections of which are (generalized) ``matter fields''.  Likewise, matter fields may be associated with other generalized fields on spacetime, again construed as sections of vector bundles, representing properties matter may have, such as velocity, electromagnetic charge, color charge, isospin, etc.  Both the states of matter and their associated properties may be construed as geometrical, not because they have some special relationship to space and time (though some, such as velocity and spin, do), but rather because they can be represented at a point by mathematical objects with lengths and relative angles.  Most importantly, according to Yang-Mills theory, the bundles in which these fields are valued are curved, where the curvature depends on the distribution of the corresponding properties throughout space and time.  The effects that we might have otherwise thought of as due to the electromagnetic, weak, or strong forces are now understood as manifestations of this curvature.

Of course, putting Yang-Mills theory in these informal terms makes the analogy with general relativity especially striking.  But this is not to say that the theories are the same.  As I discussed in the last section, there are significant disanalogies between Yang-Mills theory and general relativity.  Nonetheless, I have argued that these disanalogies do not bear on the picture of Yang-Mills theory just sketched.  The reason is that ultimately, the only analogy the present argument requires is that one can think of the principal bundles one encounters in both general relativity and in Yang-Mills theory as (subbundles of) frame bundles, and thus interpret sections of principal bundles in both cases as frame fields for vector bundles.  That one \emph{should} interpret the frame bundle in general relativity in this way seems uncontentious; that one may also interpret the principal bundles in Yang-Mills theory in the same way is perhaps less obvious, but nonetheless viable---and, it seems to me, attractive.

\appendix

\section{Primer on the Geometry of Yang-Mills Theory for Philosophers of Space and Time}\label{Preliminaries}

In this appendix, I provide background definitions and some further technical details regarding principal bundles and vector bundles.  Unlike other presentations of this material, this appendix is targeted at philosophers of physics familiar with the formalism and conventions of the foundations of spacetime physics literature, as in (for instance) \citet{Friedman}, \citet{Wald}, \citet{Earman}, or \citet{MalamentGR}.  That said, this is not a complete pedagogical treatment of these topics.  For further details, the books I find clearest are \citet{Palais} and \citet{Bleecker}; \citet{Nakahara} and \citet{Baez+Munian} are also noteworthy.  For additional background on the geometry, the classic source on the theory of connections on fiber bundles is \citet{Kobayashi+Nomizu}, which remains the most comprehensive text available, despite certain pre-modern tendencies; additional helpful sources are \citet{Kolar+etal}, \citet{Michor}, \citet{Lee2}, and \citet{Taubes}.  One important difference, however, between the presentation here and in the sources just cited is that, as noted in fn. \ref{AI}, I use the ``abstract index'' notation developed by \citet{Penrose+Rindler} and described in detail by \citet{Wald} and \citet{MalamentGR}, suitably modified to distinguish the range of vector spaces that one encounters in the theory of principal bundles.\footnote{See \citet{Geroch} and \citet[Ch. 13]{Wald} for similar generalizations of the notation.  The principal virtues to using this notation, aside from familiarity to a particular community, is that it naturally permits multi-index tensor fields, as discussed in section \ref{AppNotation} below.}

\subsection{Fiber bundles}\label{AppFiberBundles}
A \emph{(smooth)}\footnote{In what follows, as in the body of the paper, it should be assumed that all manifolds, maps, and fields described are smooth.  Likewise, all manifolds are assumed to be Hausdorff and paracompact.} \emph{fiber bundle} $B\xrightarrow{\pi} M$ is a smooth, surjective map $\pi$ from a manifold $B$ to a manifold $M$ satisfying the following condition: there exists a manifold $F$ such that, given any point $p\in M$, there exists an open neighborhood $U\subseteq M$ containing $p$ and a \emph{local trivialization of $B$ over $U$}, which is a diffeomorphism $\zeta: U\times F\rightarrow \pi^{-1}[U]$ such that $\pi\circ\zeta:(q,f)\mapsto q$ for all $(q,f)\in U\times F$.  The map $\pi$ is called the \emph{projection map}; the manifold $B$ is called the \emph{total space} of the bundle; the manifold $M$ is called the \emph{base space}; and the manifold $F$ is called the \emph{typical fiber}.  The local trivialization condition guarantees that the collection of points in $B$ mapped by $\pi$ to a given point $p\in M$, denoted $\pi^{-1}[p]$ and called the \emph{fiber at $p$}, is an embedded submanifold of the total space diffeomorphic to the typical fiber.  This fact supports the following picture: a fiber bundle may be thought of as an association of copies of $F$ with each point of $M$ in such a way that the result is ``locally a product manifold'' in much the same way that a manifold is ``locally $\mathbb{R}^n$''.  We will sometimes write $F\rightarrow B\xrightarrow{\pi} M$ when we want to emphasize the typical fiber of a given fiber bundle; under other circumstances, when no ambiguity can arise, we will use just the total space $B$ to refer to the entire bundle.

A \emph{fiber bundle morphism} $(\Psi,\psi):(B\xrightarrow{\pi} M)\rightarrow (B'\xrightarrow{\pi'} M)$ is a pair of smooth maps $\Psi:B \rightarrow B'$ and $\psi:M\rightarrow M'$ such that $\pi'\circ\Psi=\psi\circ\pi$. A fiber bundle morphism is said to be an \emph{isomorphism} if both maps are diffeomorphisms.  A fiber bundle isomorphism whose domain and codomain are the same (i.e., a fiber bundle automorphism) is said to be \emph{vertical} if the associated map $\psi:M\rightarrow M$ is the identity---i.e., if the automorphism takes the fiber over each point back to itself.  A \emph{(local) section} of a fiber bundle $B\xrightarrow{\pi} M$ is a smooth map $\sigma: U\rightarrow B$ such that $\pi\circ\sigma = 1_U$, where $U\subseteq M$ is open and $1_U$ is the identity on $M$ restricted to $U$. A local section of a fiber bundle may be thought of as a generalization of the ordinary notion of smooth (scalar, vector, tensor) ``field'' on a manifold: it is a smoothly-varying assignment of a fiber value to each point $p\in U$.

Examples of fiber bundles include product manifolds with the projection onto one of the factors as the projection map, such as $N\rightarrow M\times N\xrightarrow{pr_1} M$.  A fiber bundle that can be written this way, i.e., any bundle admitting a \emph{global trivialization}, is called a \emph{trivial bundle}.\footnote{Note that a trivial bundle and a product manifold are not quite the same thing: a fiber bundle only has one privileged projection map, which is onto the base space, whereas a product manifold has two privileged projection maps, one onto each factor.}  An example of a non-trivial fiber bundle is the M\"obius strip, $(-1,1)\rightarrow M\ddot{o}\xrightarrow{\pi} S^1$, which has the circle as base space and an open subset of the real line as fiber.  Here the fibers are ``twisted'' in such a way that $M\ddot{o}$ is not isomorphic to the cylinder, $S^1\times (-1,1)$.

\subsection{Vector bundles and principal bundles}\label{AppVector+PrincipalBundles}
We will be particularly interested in two special classes of fiber bundles: principal bundles and vector bundles.   A \emph{vector bundle} is a fiber bundle $V\rightarrow E \xrightarrow{\pi} M$ where the typical fiber $V$ is a vector space and for each $p\in M$, there exists a neighborhood $U$ of $p$ and a local trivialization $\zeta:U\times V\rightarrow \pi^{-1}[U]$ such that for any $q\in U$, the map $v\mapsto \zeta(q,v)$ is a vector space isomorphism.  A smooth fiber bundle morphism $(\Psi,\psi):(E\xrightarrow{\pi} M)\rightarrow(E'\xrightarrow{\pi'} M')$ is a \emph{vector bundle morphism} if for each $p\in M$, the restricted map $\Psi_{|\pi^{-1}[p]}:\pi^{-1}[p]\rightarrow \pi'^{-1}[\psi(p)]$ is linear; it is a \emph{vector bundle isomorphism} if it is also a fiber bundle isomorphism.

A \emph{principal bundle}, meanwhile, is a fiber bundle $G\rightarrow P\xrightarrow{\wp} M$ where $G$ is a Lie group---i.e., a smooth manifold endowed with a group structure in such a way that the group operations are smooth maps---and there is a smooth, free,\footnote{A right action of a group $G$ on a space $P$ is \emph{free} if for any point $x\in P$ and $g\in G$, $xg=x$ if and only if $g$ is the identity.}  fiber-preserving right action of $G$ on $P$ such that given any point $p\in M$, there exists a neighborhood $U$ of $p$ and a local trivialization $\zeta:U\times G\rightarrow \wp^{-1}[U]$ such that for any $q\in U$ and any $g,g'\in G$, $\zeta(q,g)g'=\zeta(q,gg')$.  The group $G$ is known as the \emph{structure group} of the bundle.  A \emph{principal bundle morphism} consists of a fiber bundle morphism $(\Psi,\psi):(G\rightarrow P\xrightarrow{\wp} M)\rightarrow (G'\rightarrow P'\xrightarrow{\wp'} M')$ and a smooth homomorphism $h:G\rightarrow G'$ such that for any $x\in P$ and $g\in G$, $\Psi(xg)=\Psi(x)h(g)$.  A principal bundle morphism is a \emph{principal bundle isomorphism} if $(\Psi,\psi)$ and $h$ are isomorphisms; it is a \emph{vertical principal bundle automorphism} if $(\Psi,\psi)$ is a vertical bundle automorphism and $h$ is the identity map.

%There are clear analogies between our definitions of vector bundles and principal bundles, but there is one crucial disanalogy worth remarking on.  The fiber at each point of a vector bundle has a distinguished, trivialization-independent vector space structure, and hence there is a privileged $0$ element in each fiber, one can add fiber elements, etc.  The fiber at each point of a principal bundle, in contrast, does \emph{not} have a fixed group structure.  There is no distinguished identity element or group multiplication, except relative to a choice of local trivialization.  Note, too, that given a principal bundle $G\rightarrow P\xrightarrow{\wp} M$, the right action of a group element $g\in G$, $x\mapsto xg$, does not generally induce a principal bundle automorphism, since given any other element $g'\in G$, $xg'\mapsto xg'g\neq xgg'$.  In fact, the right action by an element $g\in G$ yields an automorphism just in case $g$ is in the center of $G$ (i.e., it commutes with every element of $G$).

\subsection{Notational Conventions}\label{AppNotation}
Let $F\rightarrow B\xrightarrow{\pi}M$ be a fiber bundle.  I will adopt the following notational conventions.\footnote{As noted above, these should be understood in the context of the abstract index notation.  The particular conventions introduced here follow \citet{Geroch} closely.}  I will use lower-case Latin indices $a,b,c,\ldots$ for vectors and tensors that are tangent to the base space of a bundle, or generically when I am considering manifolds outside the context of a particular bundle.  Lower-case Greek indices $\alpha,\beta,\gamma\ldots$ will label vectors and tensors that are tangent to the total space of a bundle.  So given a point $x\in B$, a vector at $x$ would be denoted (for instance) by $\xi^{\alpha}$, while a vector at $\pi(x)$ would be denoted by $\eta^a$.  Capital Latin indices $A,B,C,\ldots$ will label vectors and tensors valued in other spaces, including the fibers of vector bundles.  In cases where there are several such vector spaces under consideration, further decorations will be used to distinguish membership in the different spaces. Finally, if a given vector space has a Lie algebra structure, we will label vectors in that space using capital Fraktur indices $\mathfrak{A},\mathfrak{B},\mathfrak{C},\ldots$. In all cases, raised indices will indicate that an object is an element of a salient vector space; lowered indices will indicate that the object is a linear functional on the corresponding vector space.

This notation allows one to consider ``mixed index'' tensors and tensor fields on a manifold $M$, which represent multilinear maps between various vector spaces associated with a point of $M$.  For instance, given a vector $\xi^{\alpha}$ at a point $x$ in the total space of a fiber bundle, one may think of the pushforward along $\pi$ at $x$, $(\pi_x)_*$, as a linear map from vectors at $x$ to vectors at $\pi(x)$, which might then be written as $(\nabla\pi)^a{}_{\alpha}$.  In fact, this notation also subsumes the pullback along $\pi$, since given a covector $\kappa_a$ at $\pi(x)$, $\kappa_a(\nabla\pi)^a{}_{\alpha}$ is precisely the covector at $x$ whose action on a vector $\xi^{\alpha}$ is the action of $\kappa_a$ on the pushforward of $\xi^{\alpha}$, i.e., $(\kappa_a(\nabla\pi)^a{}_{\alpha})\xi^{\alpha}=\kappa_a((\nabla\pi)^a{}_{\alpha}\xi^{\alpha})$.  In what follows, I will freely adopt both the mixed index and $*$ notations for the pushforward and pullback, depending on context.  Other examples of mixed index tensors include principal connections (discussed in \ref{AppPrincipalConnections}), curvature tensors (discussed in \ref{AppCurvature}), and solder forms (discussed in section \ref{Disanalogies}).  Mixed index tensors on a manifold $M$ are said to be smooth if their contraction with appropriate collections of smooth vectors (and covectors) is a smooth scalar field.\footnote{See \ref{AppVectorValuedForms} for a degenerate instance of this criterion that may clarify how it works in practice.}

\subsection{Tangent and Frame Bundles}\label{AppTangent+FrameBundles}
Any manifold $M$ is naturally associated with a vector bundle over $M$, known as the \emph{tangent bundle}. Let $T_pM$ be the tangent space at $p\in M$ and let $TM$ be the set of all of the tangent vectors at all point of $M$, $TM= \bigcup_{p\in M} T_pM$.  A manifold structure on $TM$ may be induced as follows.  First note that any point $x\in TM$ may be written as $(p,\xi^a)$, where $\xi^a$ is a tangent vector at $p$.  Then, given any chart $(U,\varphi)$ on $M$, one can associate a point $(p,\xi^a)\in TM$ with an element of $\mathbb{R}^{2n}$ (where $n$ is the dimension of $M$) by $(p,\xi^a)\mapsto (\varphi^1(p),\ldots,\varphi^n(p),\overset{1}{\xi},\ldots \overset{n}{\xi})$.  Here $\varphi^i(p)$ represents the $i$th coordinate of $p$ relative to the chosen chart, and $\overset{i}{\xi}$ is the $i$th component of $\xi^a$ in the basis determined by the chart.  Requiring all such maps, for every chart on $M$, to be smooth and smoothly invertible induces a manifold structure on $TM$.  The map $\pi_T:TM\rightarrow M$ that takes elements $x\in TM$ to the point $p\in M$ to which they correspond---that is, $\pi_T:(p,\xi^a)\mapsto p$---is smooth relative to this manifold structure.  Similarly, the maps used to induce the manifold structure on $TM$ are local trivializations relative to which $\mathbb{R}^n\rightarrow TM\xrightarrow{\pi_T} M$ is a vector bundle over $M$.  Sections of the tangent bundle are (ordinary, tangent) vector fields, i.e., smooth assignments of tangent vectors to point of (an open subset of) a manifold. Similar constructions may be used to define the \emph{cotangent bundle}, $T^*M\xrightarrow{\pi_{T^*}}M$, and bundles of rank $(r,s)$ tensors on $M$.

The manifold $M$ also naturally determines a principal bundle over $M$, known as the \emph{frame bundle}.  The construction is similar to the tangent bundle.  Suppose $n$ is the dimension of $M$.  Then a \emph{frame} at a point $p$ is an ordered collection of $n$ linearly independent vectors at $p$.  Let $LM$ be the collection of all frames at all points of $M$.  Analogously to the tangent bundle, any point $x\in LM$ may be written $(p,u)$, where $u=(\overset{1}{u}{}^a,\ldots,\overset{n}{u}{}^a)$ is a frame at $p$.  Now given a chart $(U,\varphi)$ on $M$, we may associate any point $(p,u)\in LM$ with an element of $\mathbb{R}^{n^2+n}$ by $(p,u)\mapsto (\varphi^1(p),\ldots,\varphi^n(p),\overset{11}{u},\ldots,\overset{ij}{u},\ldots,\overset{nn}{u})$, where now $\overset{ij}{u}$ is the $j$th component in the basis determined by the chart of $\overset{i}{u}{}^a$, the $i$th element of the frame.  The image of this map is $\varphi[U]\times F$, where $F\subset\mathbb{R}^{n^2}$ is the collection of all $n^2$-tuples corresponding to invertible $n\times n$ matrices.  Thus $F$ is an open subset of $\mathbb{R}^{n^2}$ (because the determinant map $det:\mathbb{R}^{n^2}\rightarrow \mathbb{R}$ is continuous and therefore the set of matrices with determinant $0$, $det^{-1}[0]$, is closed), diffeomorphic to the Lie group $GL(n,\mathbb{R})$ of invertible real valued matrices.  (The Lie algebra associated with $GL(n,\mathbb{R})$, $\mathfrak{gl}(n,\mathbb{R})$, is the algebra of all $n\times n$ matrices, not necessarily invertible.)  Requiring all such maps, for all charts on $M$, to be smooth and smoothly invertible induces a manifold structure on $LM$.  The map $\wp_L:LM\rightarrow M$ where $\wp_L:(p,u)\mapsto p$ is smooth with respect to this structure, and once again the chart-relative maps defined above are local trivializations relative to which $GL(n,\mathbb{R})\rightarrow LM\xrightarrow{\wp_L} M$ is a principal bundle, where the associated right $GL(n,\mathbb{R})$ action corresponds to a smooth change of frame at each point.  Sections of the frame bundle are (local) \emph{frame fields}, which are smoothly varying bases of the tangent space assigned to each point of (an open subset of) a manifold.

More generally, given any vector bundle $V\rightarrow E\rightarrow M$, one may readily construct an associated principal bundle $GL(V)\rightarrow LE\rightarrow M$, called the frame bundle for $E$, whose fibers correspond to the frames for the fibers of $E$.

\subsection{Associated bundles}\label{AppAssociatedBundles}
The frame bundle construction provides a sense in which a given vector bundle may be used to build a principal bundle.  But one can also move in the other direction, from a given principal bundle to a vector bundle.  Let $G\rightarrow P\xrightarrow{\wp} M$ be a principal bundle, and let $V$ be some vector space with a (fixed) representation $\rho:G\rightarrow GL(V)$ of $G$.  Now for each point $p\in M$, consider maps $v^A:\wp^{-1}[p]\rightarrow V$ that are equivariant in the sense that for any $x\in\wp^{-1}[p]$, $v^A(xg)=(\rho(g^{-1}))^A{}_B v^B$ and smooth in the sense that, given any fixed linear functional $u_A$ on $V$, $u_Av^A$ is a smooth scalar field on $\pi^{-1}[p]$.  Let $P\times_{G} V$ denote the set of all such maps, for all points $p\in M$.  Then there is a unique manifold structure on $P\times_{G} V$ such that $V\rightarrow P\times_G V\xrightarrow{\pi_{\rho}} M$, where $\pi_{\rho}:(v^A:\pi^{-1}[p]\rightarrow V)\mapsto p)$, is a vector bundle over $M$.  This bundle is called an \emph{associated vector bundle}.  Under this construction, if $V$ is an $n$ dimensional vector space, then any vector bundle $V\rightarrow E\xrightarrow{\pi} M$ is isomorphic to $LE\times_{GL(V)} V\xrightarrow{\pi_{\rho}}M$, for any faithful representation of $GL(V)$ on $V$, where $LE\xrightarrow{\wp_L} M$ is the frame bundle for $E$.

\subsection{Vector valued forms}\label{AppVectorValuedForms}
Let $M$ be a manifold and let $V$ be a vector space.  A \emph{vector valued $n$ form} (or a $V$ valued $n$ form) on an open set $U$ is a mixed index tensor field $\kappa^A{}_{a_1\cdots a_n}$ that is totally antisymmetric in its covariant (tangent) indices, and where the $A$ index indicates membership in the (fixed) vector space $V$.\footnote{By ``fixed'', I mean that all of these maps have the same vector space as their codomain, i.e., the vector space does not vary from point to point of $M$ as it would with a vector bundle over $M$.   Another, more general, way of thinking about vector valued $n$ forms is as mixed index tensors on $M$, $\kappa^A{}_{a_1\cdots a_n}(=\kappa^A{}_{[a_1\cdots a_n]})$, whose single contravariant index is valued in the fibers of some vector bundle over $M$, rather than a single vector space $V$.  But one has to be careful. If one adopts this more general perspective, one \emph{cannot} extend the exterior derivative from ordinary forms to vector valued forms except in the presence of a linear connection on the vector bundle over $M$. See \citet[pp. 10-11 \& 30-32]{Palais} for discussion.}  These fields are required to satisfy the following (degenerate) smoothness condition: given any (fixed) linear functional $\beta_A$ on $V$, we require that $\beta_A\alpha^A{}_{a_1\cdots a_n}$ is a smooth (ordinary) $n$ form on $U$.  Note that from this perspective, ordinary $n$ forms are also vector valued forms, where the vector space in which they are valued is $\mathbb{R}$; in such cases, one simply drops the index corresponding to membership in $\mathbb{R}$.

Recall that there is a natural notion of differentiation on (ordinary, $\mathbb{R}$ valued) $n$ forms on $M$, given by the exterior derivative $d_a$, which takes $n$ forms $\kappa_{a_1\cdots a_n}$ to $(n+1)$ forms $d_a\kappa_{a_1\ldots a_n}$.  The exterior derivative extends to vector valued $n$ forms as follows.  Given a (smooth) $V$ valued $n$ form $\kappa^A{}_{a_1\cdots a_n}$, we take the exterior derivative of $\kappa^A{}_{a_1\cdots a_n}$, written $d_a\kappa^A{}_{a_1\cdots a_n}$, to be the unique $V$ valued $(n+1)$ form whose action on any (fixed) linear functional $\beta_A$ on $V$ is given by $\beta_A(d_n\alpha^A{}_{a_1\cdots a_n})=d_n(\beta_A\alpha^A{}_{a_1\cdots a_n})$, where the expression on the right should be interpreted as the (ordinary) exterior derivative of the (ordinary, smooth) $n$ form $\beta_A\alpha^A{}_{a_1\cdots a_n}$.\footnote{I am grateful to Dick Palais (personal correspondence) for suggesting this way of thinking about the exterior derivative's action on vector valued forms to me.}  Similarly, given a smooth map $\varphi:M\rightarrow N$ and a $V$ valued $n$ form $\alpha^A{}_{a_1\cdots a_n}$ on $N$, one can define the pullback of $\alpha^A{}_{a_1\cdots a_n}$ along $\varphi$, $\varphi^*(\alpha^A{}_{a_1\cdots a_n})$, as the unique $V$ valued $n$ form on $M$ such that, given any (fixed) linear functional $\beta_A$ acting on $V$, $\beta_A\varphi^*(\alpha^A{}_{a_1\cdots a_n})=\varphi^*(\beta_A\alpha^A{}_{a_1\cdots a_n})$.

Finally, consider the special case of vector valued $n$ forms on the total space $P$ of a principal bundle $G\rightarrow P\xrightarrow{\wp} M$ (or, more generally, forms defined on $\wp^{-1}[U]$, for some open $U\subset M$).  In this context, it is often natural to fix a representation $\rho$ of $G$, the structure group of the bundle, on the vector spaces in which forms of interest are valued.  One can then consider forms that are \emph{equivariant} with respect to the $G$ action on $P$, in the sense that, given a $V$ valued $n$ form $\kappa^A{}_{\alpha_1\cdots\alpha_n}$ on $P$, $\kappa^A{}_{\alpha_1\cdots\alpha_n}$ is such that given any element $g\in G$, any point $x\in P$, and any vectors $\overset{1}{\eta}{}^{\alpha},\ldots,\overset{n}{\eta}{}^{\alpha}$ at $x$, the following condition holds: \begin{equation}\label{equivariant}
(\kappa^{A}{}_{\alpha_1\cdots\alpha_n})_{|xg}(R_g)_*(\overset{1}{\eta}{}^{\alpha_1}\cdots\overset{n}{\eta}{}^{\alpha_n})=((\rho(g^{-1}))^{A}{}_{B}\kappa^{B}{}_{\alpha_1\cdots\alpha_n})_{|x}\overset{1}{\eta}{}^{\alpha_1}\cdots \overset{n}{\eta}{}^{\alpha_n}.\end{equation} Here $(\rho(g^{-1}))^{A}{}_{B}$ is the tensor acting on $V$ corresponding to $g^{-1}$ in the representation $\rho$, and $R_g$ is the smooth right action of $g$ on $P$.  Note that this equation makes sense, since both sides are elements of the fixed vector space $V$.  Note, too, that, following the construction of the previous section, we now see that sections $\kappa^A:U\rightarrow P\times_G V$ of the associated bundle $V\rightarrow P\times_G V\xrightarrow{\pi_{\rho}}M$ may be identified with equivariant $V$ valued $0$ forms on $\wp^{-1}[U]$.

\subsection{Connections and parallel transport}\label{AppConnections}

We now turn to connections.  Some preliminary definitions are in order.  Fix an arbitrary fiber bundle $F\rightarrow B\xrightarrow{\pi} M$ and let $\xi^{\alpha}$ be a vector at a point $x\in B$.  We will say $\xi^{\alpha}$ is \emph{vertical} if $(\nabla\pi)^a{}_{\alpha}\xi^{\alpha}=\mathbf{0}$.  Since $(\nabla\pi)^a{}_{\alpha}$ is a linear map, its kernel forms a linear subspace of $T_x B$, written $V_x$ and called the \emph{vertical subspace} at $x$; in general, the dimension of $V_x$ will be the dimension of $F$.  Indeed, one can think of the vertical vectors at $x$ as ``tangent to the fiber'' in the precise sense that they are tangents to curves through $x$ that remain in the fiber over $\pi(x)$.  If a vector $\xi^{\alpha}$ at a point $x$ is not vertical, then it is \emph{horizontal}.  A subspace $H_x$ of $T_xB$ will be called a \emph{horizontal subspace} if any vector $\xi^{\alpha}$ at $x$ may be uniquely written as the sum of one vector in $V_x$ and one vector in $H_x$.  It immediately follows that, given any horizontal subspace $H_x$ at $x$, $(\nabla\pi)^a{}_{\alpha}:H_x\rightarrow T_{\pi(x)}$ is a vector space isomorphism, and the dimension of any horizontal subspace is the same as the dimension of $M$.

Though the vertical subspace is uniquely determined by the map $\pi$, there is considerable freedom in the choice of horizontal subspace.  A connection on $B$, roughly speaking, is a smoothly varying choice of horizontal subspace at each point $x\in B$.  This idea of ``smoothly varying'' may be made precise as follows.  Given any point $x\in B$ and a horizontal subspace $H_x$, one can always find a tensor $\omega^{\alpha|}{}_{\beta}$ that acts on any vector $\xi^{\alpha}$ at $x$ by projecting $\xi^{\alpha}$ onto its (unique, relative to $H_x$) vertical component, $\omega^{\alpha|}{}_{\beta}\xi^{\beta}$.  (Here we have used a $|$ following a Greek index to emphasize that the index is vertical.) Conversely, given any tensor $\omega^{\alpha|}{}_{\beta}$ at $x$ such that (a) $\omega^{\alpha|}{}_{\beta}\omega^{\beta|}{}_{\kappa}=\omega^{\alpha|}{}_{\kappa}$ and (b) given any vertical vector $\xi^{\alpha|}$ at $x$, $\omega^{\alpha|}{}_{\beta}\xi^{\beta|}=\xi^{\alpha|}$, one can always define a horizontal subspace $H_x$ at $x$ as the kernel of $\omega^{\alpha|}{}_{\beta}$.  This permits us to adopt the following official definition of a connection: a \emph{connection} on a fiber bundle $B\xrightarrow{\pi} M$ is a smooth tensor field $\omega^{\alpha|}{}_{\beta}$ on $B$ satisfying conditions (a) and (b) above.  Every fiber bundle admits connections.  %Note that the projection map $\pi$ and a connection $\omega^{\alpha'}{}_{\beta}$ together yield an isomorphism between the horizontal spaces $H_x$ at any $x\in B$ and the tangent space of the manifold point below $x$, $\pi(x)$, which may be represented by a mixed index tensor field $\hat{\omega}^{\alpha}{}_a$ on $B$ with the properties that (1) $\omega^{\alpha|}{}_{\beta}\hat{\omega}^{\beta}{}_a=0$ and $(\nabla\pi)^a{}_{\alpha}\hat{\omega}^{\alpha}{}_b=\delta^a{}_b$.  One may think of $(\nabla\pi)^a{}_{\alpha}$, restricted to horizontal vectors, as the inverse of $\hat{\omega}^{\alpha}{}_a$.

A connection provides a notion of parallel transport of fiber values along curves in the base space $M$.  This works as follows.  Consider a smooth curve $\gamma:I\rightarrow M$.  One can always \emph{lift} such a curve $\gamma$ into the total space $B$, by defining a new curve $\hat{\gamma}:I\rightarrow B$ with the property that $\pi\circ\hat{\gamma}=\gamma$.  This curve $\hat{\gamma}$ is generally not unique.  One gets a unique lift by specifying some additional data: fix a connection $\omega^{\alpha|}{}_{\beta}$ and choose some $t_0\in I$ and some $x\in\pi^{-1}[\gamma(t_0)]$---that is, choose some point in the fiber above $\gamma(t_0)$.  Then there is a unique \emph{horizontal lift} of $\gamma$ through $x$, that is, a unique lift $\hat{\gamma}:I\rightarrow E$ such that (1) $\hat{\gamma}(t_0)=x$ and (2) the vector tangent to $\hat{\gamma}$ at each point in its image is horizontal relative to $\omega^{\alpha|}{}_{\beta}$.  Then, given any $t\in I$, we say the \emph{parallel transport} of $x$ to $\gamma(t)$ along $\gamma$ is $\hat{\gamma}(t)$, which is a point in the fiber above $\gamma(t)$.

\subsection{Principal connections}\label{AppPrincipalConnections}
Now suppose one has a principal bundle $G\rightarrow P\xrightarrow{\wp} M$.  Here, one is interested in connections that are compatible with the principal bundle structure in the sense that they are equivariant under the right action of $G$ on $P$.  A bit of work is required to make this precise.  First, recall that given any Lie group $G$, $T_eG$, the tangent space at the identity element, is endowed with a natural Lie algebra structure induced from the Lie group structure as follows.  Take any vector $\xi^a\in T_eG$.  One can define a (smooth) vector field on $G$, called a \emph{left-invariant vector field} by assigning to each point $g\in G$ the vector $({}^g\ell_e)_*(\xi^a)$, i.e., the pushforward of $\xi^a$ along the left action on $G$ determined by $g$.  The Lie bracket of two vectors at $e$, then, is just the ordinary commutator of the corresponding vector fields induced by the left action.\footnote{Given a manifold $M$, a point $p$, and two (tangent) vector fields $\xi^a$ and $\eta^a$ defined on some neighborhood $O$ containing $p$, the \emph{commutator} of $\xi^a$ and $\eta^a$, written $[\xi,\eta]^a$ or $[\xi^a,\eta^a]$, is defined by $[\xi,\eta]^a=\mathcal{L}_{\xi}\eta^a$, where $\mathcal{L}_{\xi}$ is the Lie derivative with respect to $\xi^a$, defined in section \ref{Disanalogies}.  For more on the Lie derivative and commutator, see \citet[\S 1.6]{MalamentGR}.} The Lie algebra associated in this way with a Lie group $G$ is often denoted $\mathfrak{g}$.  The Lie algebra $\mathfrak{g}$, understood as a vector space, comes with a privileged representation of $G$, called the \emph{adjoint representation}, $ad:G\rightarrow GL(\mathfrak{g})$, defined by $(ad(g))^{\mathfrak{A}}{}_{\mathfrak{B}}=(\nabla\Upsilon^g_{|e})^{\mathfrak{A}}{}_{\mathfrak{B}}$, where (1) $\Upsilon^g:G\rightarrow G$ acts as $\Upsilon^g:h\mapsto ghg^{-1}$, and (2) $(\nabla\Upsilon^g_{|e})^{\mathfrak{A}}{}_{\mathfrak{B}}$ should be understood as the pushforward along $\Upsilon^g$ at the identity, which maps $T_eG$ to itself because $\Upsilon^g(e)=geg^{-1}=e$ for every $g\in G$.

The right action of the structure group $G$ on the principal bundle $P$ allows us to define a canonical isomorphism between the vertical space $V_x$ at each point $x\in P$ and the Lie algebra $\mathfrak{g}$ associated with $G$.  Given any vector $\xi^{\mathfrak{A}}\in T_eG$, let $\gamma_{\xi}:I\rightarrow G$ be the (sufficiently unique) integral curve of the left-invariant vector field associated with $\xi^{\mathfrak{A}}$.  We assume $\gamma(0)=e$.  Then, given any point $x\in P$, one can define a curve $\tilde{\gamma}_{\xi}:I\rightarrow P$ through $x$ by setting $\tilde{\gamma}_{\xi}(t)=x\gamma_{\xi}(t)$.  We will take the tangent to this curve, $\overrightarrow{(\tilde{\gamma}_{\xi})}^{\alpha|}$, which is necessarily vertical because the right action of $G$ on $P$ is fiber-preserving, to be the vertical vector at $x$ associated with $\xi^{\mathfrak{A}}$.  This construction defines a linear bijection that can be represented by a mixed index tensor field on $P$, $\mathfrak{g}^{\mathfrak{A}}{}_{\alpha|}$, with inverse $\mathfrak{g}^{\alpha|}{}_{\mathfrak{A}}$.  (Here the $|$ next to a covariant index means that $\mathfrak{g}^{\mathfrak{A}}{}_{\alpha |}$ is only defined for vertical vectors.)  Thus, we can always think of vertical vectors at a point $x$ of a principal bundle as elements of a \emph{fixed} Lie algebra, independent of $x$.

The isomorphism just described allows one to think of a connection $\omega^{\alpha|}{}_{\beta}$ on a principal bundle as a vector valued one form (sometimes called a \emph{Lie algebra valued one form}), given by $\omega^{\mathfrak{A}}{}_{\beta}=\mathfrak{g}^{\mathfrak{A}}{}_{\alpha|}\omega^{\alpha|}{}_{\beta}$.  Since this is a vector valued form on $P$, we also fix a representation of $G$ on $\mathfrak{g}$, the vector space in which the form is valued (recall \ref{AppVectorValuedForms}); here, as with all Lie algebra valued forms we will consider, we take $G$ to act on $\mathfrak{g}$ in the adjoint representation.  Finally, we may define a \emph{principal connection} as a connection $\omega^{\mathfrak{A}}{}_{\alpha}$ on $P$ that is equivariant in the sense of Eq. \eqref{equivariant}.  This condition guarantees that for any point $x\in P$ and any $g\in G$, the horizontal subspace determined by $\omega^{\mathfrak{A}}{}_{\beta}$ at $xg$ equals the pushforward of the horizontal space at $x$ along the right action of $g$, i.e., $({}^gR)_*[H_x]=H_{xg}$.  Every principal bundle admits principal connections.

\subsection{Covariant Derivatives}\label{AppCovariantDerivatives}
Consider a vector bundle $V\rightarrow E\xrightarrow{\pi} M$.   A \emph{covariant derivative operator} $\nabla$ on $E$ is a map from sections $\sigma^A:U\rightarrow E$ of $E$ to mixed index tensor fields $\sigma^A\mapsto \nabla_a\sigma^A$ on $U$ satisfying the following conditions: (1) given two smooth sections $\sigma^A,\nu^A:U\rightarrow M$, $\nabla_a(\sigma^A + \nu^A)=\nabla_a\sigma^A + \nabla_a\nu^A$ and (2) given any smooth scalar field $\lambda:M\rightarrow \mathbb{R}$, $\nabla_a (\lambda v^A)=v^A d_a\lambda + \lambda \nabla_a v^A$, where $d_a$ is the exterior derivative.  The covariant derivative of a section $\sigma^A:U\rightarrow E$ has the following interpretation.  Given a vector $\xi^a$ at a point $p\in U$, $\xi^a\nabla_a\sigma^A$ is the derivative of $\sigma^A$ in the direction of $\xi^a$, relative to a standard of fiber-to-fiber constancy given by $\nabla$. The covariant derivatives one encounters in general relativity are special cases of this more general definition, where the vector bundle in question is the tangent bundle (and, by extension, various bundles of tensors constructed out of tangent vectors).  Note that a covariant derivative operator on an arbitrary vector bundle also provides a notion of parallel transport, by a construction directly analogous to that for covariant derivatives on the tangent bundle.\footnote{The construction is regrettably complicated; see \citet[\S1.7]{MalamentGR} for a detailed discussion.  Note, too, that a covariant derivative on a vector bundle is closely related to a connection on that bundle, in the sense described in \ref{AppConnections}.  Given a covariant derivative on a vector bundle, there is always a unique connection on the bundle that gives rise to the same standard of parallel transport; conversely, given any connection whose associated parallel transport generates linear maps between fibers (a so-called \emph{linear connection}), there is a unique covariant derivative operator that gives rise to the same standard of parallel transport.}

\subsection{Exterior and induced covariant derivatives}\label{AppExterior+Induced}
Now consider a principal bundle $G\rightarrow P\xrightarrow{\wp} M$ endowed with a principal connection $\omega^{\mathfrak{A}}{}_{\alpha}$ and a vector space $V$ with a fixed representation $\rho:G\rightarrow GL(V)$ of the structure group of $P$ on $V$.  In this case, we can define a second notion of differentiation for $V$ valued $n$ forms, called the \emph{exterior covariant derivative} relative to $\omega^{\mathfrak{A}}{}_{\alpha}$.  It is denoted by $\overset{\omega}{D}$.  The action of $\overset{\omega}{D}$ on a $V$ valued $n$ form $\kappa_{\alpha_1\cdots\alpha_n}$ on $U\subseteq P$ is given by $\overset{\omega}{D}_{\alpha}\kappa^A{}_{\alpha_1\cdots\alpha_n}=(d_{\beta}\kappa^A{}_{\beta_1\cdots\beta_n})\bar{\omega}{}^{\beta}{}_{\alpha}\bar{\omega}{}^{\beta_1}{}_{\alpha_1}\cdots\bar{\omega}{}^{\beta_n}{}_{\alpha_n}$, where $d$ is the ordinary exterior derivative and where $\bar{\omega}^{\alpha}{}_{\beta}=\delta^{\alpha}{}_{\beta}-\mathfrak{g}^{\alpha|}{}_{\mathfrak{A}}\omega^{\mathfrak{A}}{}_{\beta}$ is the horizontal projection relative to $\omega^{\mathfrak{A}}{}_{\alpha}$. (Recall \ref{AppPrincipalConnections}.)

A $V$ valued $n$ form $\kappa^A{}_{\alpha_1\cdots\alpha_n}$ on $\wp^{-1}[U]$, for some open $U\subseteq M$, is said to be \emph{horizontal and equivariant} if (1) it is horizontal in the sense that given any vertical vector $\xi^{\alpha}$ at a point $p\in U$, $\kappa^A{}_{\alpha_1\cdots\alpha_i\cdots \alpha_n}\xi^{\alpha_i}=\mathbf{0}$ for all $i=1,\ldots, n$ and (2) it is equivariant in the sense of Eq. \eqref{equivariant} .  The important feature of the exterior covariant derivative is that if a $V$ valued $n$ form $\kappa^A{}_{\alpha_1\cdots\alpha_n}$ is horizontal and equivariant, so is its exterior covariant derivative, $\overset{\omega}{D}_{\alpha}\kappa^A{}_{\alpha_1\cdots\alpha_n}$.  In the special case where $V$ is the Lie algebra associated with the principal bundle and $\kappa^{\mathfrak{A}}{}_{\alpha_1\cdots \alpha_n}$ is a horizontal and equivariant Lie algebra valued $n$ form, we have the relation $\overset{\omega}{D}_{\alpha}\kappa^{\mathfrak{A}}{}_{\alpha_1\cdots\alpha_n}=d_{\alpha}\kappa^{\mathfrak{A}}{}_{\alpha_1\cdots\alpha_n}+[\omega^{\mathfrak{A}}{}_{\alpha},\kappa^{\mathfrak{A}}{}_{\alpha_1\cdots\alpha_n}]$, where the bracket is the Lie bracket.

Finally, recall that in \ref{AppVectorValuedForms}, we observed that sections $\kappa^A:U\rightarrow P\times_G V$ of the associated vector bundle $V\rightarrow P\times_G V\xrightarrow{\pi_{\rho}} M$ determined by $P$, $V$, and $\rho$ are naturally understood as $V$ valued $0$ forms on (subsets of) $P$.  We now see that in fact they are horizontal and equivariant $0$ forms.  We may thus define an \emph{induced covariant derivative operator} $\overset{\omega}{\nabla}$ on $P\times_G V$ as follows: given any section $\kappa^A:U\rightarrow P\times_G V$ of $P\times_G V$, we take $\overset{\omega}{\nabla}_a\kappa^A$ to be the unique mixed index tensor on $U$ with the property that, given any point $p\in U$ and any vector $\xi^a$ at $p$, $\xi^a\overset{\omega}{\nabla}_a\kappa^A$ is the vector in the fiber of $P\times_G V$ over $p$ corresponding to the equivariant $V$ valued $0$ form $\xi^{\alpha}\overset{\omega}{D}_{\alpha}\kappa^A$ defined on the fiber of $P$ over $p$, where $\xi^{\alpha}$ is any vector field on the fiber with the property that at every point $x\in\wp^{-1}[p]$, $(\nabla\wp)^a{}_{\alpha}\xi^{\alpha}=\xi^a$.\footnote{In general many vectors at $x$ will have this property; the reason it does not matter which one chooses is that the exterior covariant derivative only acts on the horizontal part of vectors, relative to $\omega^{\mathfrak{A}}{}_{\alpha}$, and all vectors at $x$ that project down to a given vector at $\wp(x)=p$ have the same horizontal part.}  Note that this fully determines the action of $\overset{\omega}{\nabla}$ on sections of $P\times_G V$, and that, with this definition, $\overset{\omega}{\nabla}$ is both additive and satisfies the Leibniz rule (recall \ref{AppCovariantDerivatives}).  Conversely, given a principal connection on a principal bundle, this construction yields a unique covariant derivative operator on every associated vector bundle, and given a covariant derivative operator on a vector bundle, there is always a unique principal connection on the frame bundle of that vector bundle that induces the covariant derivative in this way.

\subsection{Curvature}\label{AppCurvature}
In general, the standards of parallel transport given by a principal connection or a covariant derivative operator are path-dependent.  The degree of path-dependence is measured by the \emph{curvature} of a connection or derivative operator.  Given a principal bundle $G\rightarrow P\xrightarrow{\wp} M$ and a principal connection $\omega^{\mathfrak{A}}{}_{\alpha}$ on $P$, the curvature of $\omega^{\mathfrak{A}}{}_{\alpha}$ is a horizontal and equivariant Lie algebra valued two form $\Omega^{\mathfrak{A}}{}_{\alpha\beta}$ on $P$, defined by
\begin{equation}\label{PrinCurvature}
\Omega^{\mathfrak{A}}{}_{\alpha\beta} = \overset{\omega}{D}_{\alpha}\omega^{\mathfrak{A}}{}_{\beta}.
\end{equation}
Its interpretation is as follows.  Given a point $p\in M$, an infinitesimal closed curve through $p$ may be represented by a pair of vectors, $\xi^a$ and $\eta^a$, at $p$, corresponding to the ``incoming'' and ``outgoing'' directions of the curve at $p$.  Given an arbitrary point $x\in\wp^{-1}[p]$, the (infinitesimal, or limiting) parallel transport of $x$ along this infinitesimal curve in $M$ is encoded by the vertical vector $\mathfrak{g}^{\kappa|}{}_{\mathfrak{A}}\Omega^{\mathfrak{A}}{}_{\alpha\beta}\xi^{\alpha}\eta^{\beta}$ at $x$, where $\xi^{\alpha}$ and $\eta^{\beta}$ are arbitrary vectors at $x$ with the properties that $(\nabla\wp)^a{}_{\alpha}\xi^{\alpha}=\xi^a$ and $(\nabla\wp)^a{}_{\alpha}\eta^{\alpha}=\eta^a$, respectively.  This vertical vector represents the direction and magnitude of displacement of $x$ under the infinitesimal parallel transport.

It is often convenient to express this curvature in a slightly different form, using the so-called ``structure equation'' \citep[p. 37]{Bleecker}:
\begin{equation}
\Omega^{\mathfrak{A}}{}_{\alpha\beta} = d_{\alpha}\omega^{\mathfrak{A}}{}_{\beta} + \frac{1}{2} [\omega^{\mathfrak{A}}{}_{\alpha},\omega^{\mathfrak{A}}{}_{\beta}].
\end{equation}
Here the bracket $[\cdot,\cdot]$ is the Lie bracket on the Lie algebra $\mathfrak{g}$.  It is in this form that the curvature appears in section \ref{Yang-Mills}.  (Observe that this relation is a special case of the general fact concerning exterior covariant derivatives of Lie algebra valued forms stated in \ref{AppExterior+Induced}.)

Now suppose one has a vector bundle $V\rightarrow E\xrightarrow{\pi} M$ with a covariant derivative $\nabla$ on $E$.  In this case, we may define the curvature tensor $R^A{}_{Bcd}$ as the unique  mixed index tensor on $M$ such that, given any section $\kappa^A:U\rightarrow E$, the action of $R^A{}_{Bcd}$ on $\kappa^A$ at any point $p\in U$ is:
\begin{equation}\label{VecCurvature}
R^A{}_{Bcd}\kappa^B=-2\nabla_{[c}\nabla_{d]}\kappa^A.\footnote{As with the Riemann curvature tensor, the right hand side of this equation is independent of the values of $\kappa^A$ away from $p$.  See \citet[\S1.8]{MalamentGR}.}
\end{equation}
Note that in the special case where the vector bundle is the tangent bundle, this curvature tensor corresponds exactly to the Riemann tensor.  To see the relationship between the  curvature tensors defined in Eqs. \eqref{PrinCurvature} and \eqref{VecCurvature} more clearly, note that $R^A{}_{Bcd}$ may be thought of as a mixed index tensor $\Omega^A{}_{\alpha\beta}$ on $E$, whose action at a point $v^A$ of $E$ on vectors $\xi^{\alpha},\eta^{\alpha}$ at $v^A$ is given by $(\Omega^A{}_{\alpha\beta}\xi^{\alpha}\eta^{\beta})_{|v^A}=R^A{}_{Bcd}v^B(\nabla\pi)^c{}_{\alpha}\xi^{\alpha}(\nabla\pi)^d{}_{\beta}\eta^{\beta}$.  In this form, the interpretation given above for $\Omega^{\mathfrak{A}}{}_{\alpha\beta}$ carries over essentially unchanged.\footnote{To see this, observe that the raised index on $\Omega^A{}_{\alpha\beta}$ may be thought of as ``vertical valued'', since the vertical space at any point $v^A$ of a vector bundle is canonically isomorphic to the fiber containing $v^A$.}

\subsection{Hodge star operation on horizontal and equivariant vector valued forms}\label{AppHodgeStar}
Now suppose we have a principal bundle $G\rightarrow P\xrightarrow{\wp} M$, a principal connection $\omega^{\mathfrak{A}}{}_{\alpha}$ on $P$, a metric $g_{ab}$ on $M$, and a volume element $\epsilon_{a_1\cdots a_n}$ on $M$.\footnote{For more on volume elements, see \citet[\S 1.11]{MalamentGR}.}  (Here we assume $M$ is $n$ dimensional.)  Then, given any vector space $V$, we may define a Hodge star operation, $\star$, on horizontal and equivariant $V$ valued forms on $P$.  First, note that the volume form $\epsilon_{a_1\cdots a_n}$ on $M$ determines an $n$ form $\hat{\epsilon}_{\alpha_1\cdots\alpha_n}=\wp^*(\epsilon_{a_1\cdots a_n})=\epsilon_{a_1\cdots a_n}(\nabla\wp)^{a_1}{}_{\alpha_1}\cdots(\nabla\wp)^{a_n}{}_{\alpha_n}$ on $P$.  (Observe that although $\hat{\epsilon}_{\alpha_1\cdots\alpha_n}$ is not a volume form on $P$, it is of maximal rank in the sense that any horizontal and equivariant form on $P$ with rank greater than $n$ vanishes.) Next, we define a smooth mixed index tensor field on $P$, $\bar{\omega}^{\alpha}{}_{a}$, which is uniquely characterized by the properties that (1) $(\nabla\wp)^a{}_{\alpha}\bar{\omega}^{\alpha}{}_{b}=\delta^a{}_b$ and (2) $\omega^{\mathfrak{A}}{}_{\alpha}\bar{\omega}^{\alpha}{}_a=\mathbf{0}$.  At any point $x\in P$, this tensor maps vectors $\xi^a$ at $\wp(x)$ to the unique horizontal vector $\xi^{\alpha}=\bar{\omega}{}^{\alpha}{}_a\xi^a$ at $x$ satisfying $(\nabla\wp)^a{}_{\alpha}\xi^{\alpha}=\xi^a$.  Then, for any $k\leq n$, we may define a tensor field $\hat{\epsilon}_{\alpha_1\cdots\alpha_{n-k}}{}^{\beta_1\cdots\beta_k}=\epsilon_{a_1\cdots a_{n-k}}{}^{b_1\cdots b_k}(\nabla\wp)^{a_1}{}_{\alpha_1}\cdots(\nabla\wp)^{a_{n-k}}{}_{\alpha_{n-k}}\bar{\omega}^{\beta_1}{}_{b_1}\cdots\bar{\omega}^{\beta_k}{}_{b_k}$.  (Here the indices on $\epsilon_{a_1\cdots a_n}$ are raised with $g^{ab}$.)  Finally, given any $V$ valued horizontal and equivariant $k$ form $\kappa^A{}_{\alpha_1\cdots\alpha_k}$, we may define a horizontal and equivariant $V$ valued $(n-k)$ form, $\star\kappa^A{}_{\alpha_1\cdots\alpha_{n-k}}$, by $\star\kappa^A{}_{\alpha_1\cdots\alpha_{n-k}}=\hat{\epsilon}_{\alpha_1\cdots\alpha_{n-k}}{}^{\beta_1\cdots\beta_k}\kappa^A{}_{\beta_1\cdots\beta_k}$.

Notice that this Hodge star operator has the property that, given any $V$ valued $k$ form $\kappa^A{}_{a_1\cdots a_k}$ on $M$ and any section $\sigma:U\rightarrow P$ of $P$, $\sigma^*(\star\kappa^A{}_{\alpha_1\cdots\alpha_k})=\star\sigma^*(\kappa^A{}_{\alpha_1\cdots\alpha_k})$, where the first $\star$ (acting on $\kappa^A{}_{a_1\cdots a_k}$) is the ordinary Hodge star operation, defined by $\star\kappa^A{}_{a_1\cdots a_k}=\epsilon_{a_1\cdots a_{n-k}}{}^{b_1\cdots b_k}\kappa^A{}_{b_1\cdots b_{n-k}}$.  Since acting on a horizontal and equivariant $k$ form on $P$ with the covariant exterior derivative $\overset{\omega}{D}$ yields a horizontal and equivariant $(k+1)$ form, one may always take the Hodge dual of the exterior covariant derivative of a horizontal and equivariant $k$ form $\kappa^A{}_{\alpha_1\cdots\alpha_k}$ to yield a horizontal and equivariant $(n-k-1)$ form, $\star \overset{\omega}{D}_{\alpha}\kappa^A{}_{\alpha_1\cdots\alpha_k}$.

\section*{Acknowledgments}
This material is based upon work supported by the National Science Foundation under Grant No. 1331126.  Special thanks are due to participants in my 2012 seminar on Gauge Theories, and especially Ben Feintzeig and Sarita Rosenstock for their many discussions on these topics.  I am also particularly indebted to Dick Palais and Bob Geroch for enlightening correspondences on the geometrical foundations of Yang-Mills theory.  Helpful conversations and correspondence with Dave Baker, Jeff Barrett, Gordon Belot, Erik Curiel, Katherine Brading, Richard Healey, David Malament, Oliver Pooley, Chris Smeenk, Bob Wald, David Wallace, and Chris W\"uthrich have also contributed to the development of the ideas presented here.  Erik Curiel, Sam Fletcher, and David Malament read the manuscript carefully and noted several slips (though remaining errors are, of course, my own!).  A version of this paper was presented to the Southern California Philosophy of Physics Group; I am grateful to the participants there for discussion and comments.

\singlespacing
\bibliography{yangmills}
\bibliographystyle{elsarticle-harv}
\end{document}